%% file: main.tex
\documentclass[10pt,letterpaper,compsoc,conference]{iiswc26}

\usepackage{cite}
\usepackage{amsmath,amssymb,amsfonts}
\usepackage{algorithmic}
\usepackage{graphicx}
\usepackage[table,dvipsnames]{xcolor}
\usepackage[final]{microtype}
\usepackage[italic]{mathastext}
\usepackage{libertine}
\usepackage[T1]{fontenc}
\usepackage{textcomp}
\usepackage[varqu,varl]{zi4}
\usepackage[all]{nowidow}
\usepackage[keeplastbox]{flushend}
\usepackage{fancyhdr}

\usepackage{subcaption}
\definecolor{lightgreen}{RGB}{220,245,220}
\usepackage{comment}
\usepackage{multirow}
\usepackage{xparse}
\usepackage[most]{tcolorbox}
\usepackage{dblfloatfix}
\usepackage[hidelinks]{hyperref}
\usepackage{listings}
\usepackage{enumitem}
\usepackage{xurl}

\definecolor{sqlkw}{rgb}{0.15,0.15,0.55}
\definecolor{sqlcm}{rgb}{0.40,0.40,0.40}
\definecolor{sqlbg}{rgb}{0.97,0.97,0.97}
\lstdefinestyle{sqlpaper}{%
  language=SQL,
  basicstyle=\scriptsize\ttfamily,
  keywordstyle=\color{sqlkw}\bfseries,
  commentstyle=\color{sqlcm}\itshape,
  stringstyle=\color{sqlkw},
  backgroundcolor=\color{sqlbg},
  frame=single,
  rulecolor=\color{black!25},
  framesep=2pt,
  xleftmargin=2pt,
  xrightmargin=2pt,
  showstringspaces=false,
  breaklines=true,
  columns=fullflexible,
  keepspaces=true,
  aboveskip=2pt,
  belowskip=2pt,
}
\def\BibTeX{{\rm B\kern-.05em{\sc i\kern-.025em b}\kern-.08em
    T\kern-.1667em\lower.7ex\hbox{E}\kern-.125emX}}

\newcommand{\Sirshak}[1]{}
\newcommand{\Jason}[1]{}
\newcommand{\ziyang}[1]{}
\newcommand{\daniel}[1]{}

\newcounter{observation}
\NewDocumentEnvironment{Observation}{m}{%
  \refstepcounter{observation}%
  \begin{tcolorbox}[
    colback=green!7!white,
    colframe=green!45!black,
    boxrule=0.6pt,
    arc=1.0pt,
    left=3pt,right=3pt,top=2pt,bottom=2pt,
    title={Observation~\theobservation:~#1},
    fonttitle=\bfseries,
    enhanced
  ]%
}{%
  \end{tcolorbox}%
}

\fancypagestyle{firstpage}{
  \fancyhf{}
  
}

\begin{document}
\bstctlcite{BSTcontrol}


\title{NIXT: A \underline{N}CCL \underline{I}nspector E\underline{x}porter \underline{T}ool for Observability of Collective Communication in Large Model Training}




\author{%
  \IEEEauthorblockN{Ziyang Jia\IEEEauthorrefmark{1},
                    Sirshak Das\IEEEauthorrefmark{2},
                    Jason Sewall\IEEEauthorrefmark{2},
                    Laxmi Bhuyan\IEEEauthorrefmark{1},
                    Pasha Shamis\IEEEauthorrefmark{2},
                    Daniel Wong\IEEEauthorrefmark{1}}
  \IEEEauthorblockA{\IEEEauthorrefmark{1}University of California, Riverside \qquad
                    \IEEEauthorrefmark{2}NVIDIA \\
                    \{zjia016, danwong\}@ucr.edu, bhuyan@cs.ucr.edu \qquad
                    \{sirshakd, jasewall, pshamis\}@nvidia.com}
}

\maketitle
\thispagestyle{firstpage}
\pagestyle{empty}


\begin{abstract}


As machine learning workloads scale, it is increasingly important to gain more observability into the performance of collective communication to easily identify performance variations and accelerate root cause identification. Towards this goal, the Nvidia Collective Communication Library (NCCL) introduced NCCL Inspector, a profiler plugin that provides lightweight and continuous reporting of NCCL communication performance statistics. However, the large volume of data collected by NCCL Inspector can be difficult to assess and to extract actionable insights from.

This paper presents NIXT, a \underline{N}CCL \underline{I}nspector E\underline{x}porter \underline{T}ool that improves the observability of collective communication by providing readily accessible analysis and actionable insights from NCCL Inspector profiling. To highlight the benefits of our Exporter Tool, we present a case study of Nemotron-4 LLM pretraining on an Nvidia H100 GPU cluster with up to 2,048 GPUs, demonstrate observability into how communication phases change with ML parallelism and GPU scale, and perform attribution of performance variation and root cause analysis of stragglers.

\end{abstract}


\input{1.Introduction}

\input{2.Background}

\input{3.Taxonomy}


\input{4.CaseStudy}

\input{5.NCCLtests}

\input{6.Straggler}

\input{7.Related}

\input{Conclusion}





\section*{Acknowledgments}
We would like to thank the anonymous IISWC reviewers for their invaluable comments and suggestions. We are grateful to NVIDIA for supporting this work, and huge thanks to the NVIDIA engineers and researchers whose feedback helped shape the design and evaluation of NIXT. The research was partly supported by NSF grants 2415202 and 2047521.

\bibliographystyle{IEEEtranS}
\bibliography{refs}

\clearpage
\input{artifact_appendix}

\end{document}

%% file: 1.Introduction.tex
\section{Introduction}

Collective communication~\cite{walker1996mpi,thakur2005optimization} refers to operations in which a group of GPUs cooperatively move or aggregate data. Modern distributed LLM training frameworks rely on several forms of parallelism, each of which requires a mix of collective operations, such as data~\cite{krizhevsky2014one}, tensor and model~\cite{shoeybi2019megatron}, pipeline~\cite{huang2019gpipe,10.1145/3341301.3359646}, expert~\cite{lepikhin2020gshard}, and context~\cite{liu2023ring} parallelism. 
Thus, collective communication between GPUs has become a critical performance bottleneck in large-scale clusters. To overcome this, common solutions like concurrent communication and computation \cite{10.1145/3341301.3359646} and co-designing of communication kernels and deep learning systems \cite{10.1145/3651890.3672239, 316592, 316066} have been proposed, but are still prone to performance degradation \cite{11096129}.

Distributed training spends a significant amount of time (commonly 20-40\%) on collective communication~\cite{wang2023characterizing,jia2024pccl,sergeev2018horovod}, and at large scale with thousands of GPUs, rare anomalies on individual devices (link errors, congestion, thermal throttling) can compound into job-level slowdowns and failures~\cite{erben2024hardware,cudo2024hardware,nebius2024reliable,li2024revisiting,gupta2025gpu,308748}, necessitating observability into collective communications. 
Existing tooling addresses this only partially: hardware-level GPU exporters such as DCGM~\cite{nvidia_dcgm_exporter} and Node Exporter~\cite{node_exporter} report utilization, temperature, and power but cannot attribute slowdowns to specific collectives; eBPF-based system tracing~\cite{eBPF,xu2025eacgm} sees kernel and network activity but not GPU or NCCL-level operations; framework-integrated tools such as PyTorch's NCCL Flight Recorder~\cite{pytorch_flight_recorder2024} and Mycroft~\cite{deng2025mycroft} record collective events for post-mortem debugging but are reactive and framework-specific; and hyperscale workload monitors such as FBDetect~\cite{yoon2024fbdetect} catch regressions at the job level without attributing them to individual collective operations. 
None provides a framework-agnostic, low-overhead view of collective-level behavior that can be used to characterize communication structure or perform root cause analysis.


Facing the challenge of performance observation in the large-scale training of large models (LLMs, MLLMs, VLMs), NVIDIA recently released NCCL Inspector~\cite{nvidia_nccl_inspector}, a low-overhead profiler plugin shipped with NCCL 2.28 that logs detailed performance data and metadata at runtime for every collective operation in each communicator. NCCL Inspector transparently samples ongoing collective kernels through the NCCL profiler interface and emits a structured record for every collective operation, covering message size, execution time, algorithmic and bus bandwidth, and communicator placement. Furthermore, NCCL Inspector integrates with Prometheus and Grafana dashboards for real-time performance monitoring ~\cite{nvidia_inspector_observability_2025,nvidia_inspector_prometheus_2025}.
However, the raw NCCL Inspector data is high-dimensional with thousands of operations spread across ranks, communicators, message sizes, collective types, and interconnect paths. The volume of data collected by NCCL Inspector still poses a challenge in how we can leverage this data for actionable insights into collective performance during large-scale deployment.




In order to bridge this gap, we develop NIXT, a \underline{N}CCL \underline{I}nspector E\underline{x}porter \underline{T}ool for NCCL Inspector to systematically analyze profiled data and derive meaningful insights.
To demonstrate the benefits of NIXT, we evaluate on Nemotron-4 pretraining on an H100 cluster with 16 to 2,048 GPUs as a case study to characterize collective communication in real LLM training and demonstrate straggler anomaly localization and attribution.


\begin{figure*}[!t]
  \centering
   \includegraphics[width=0.9\linewidth]{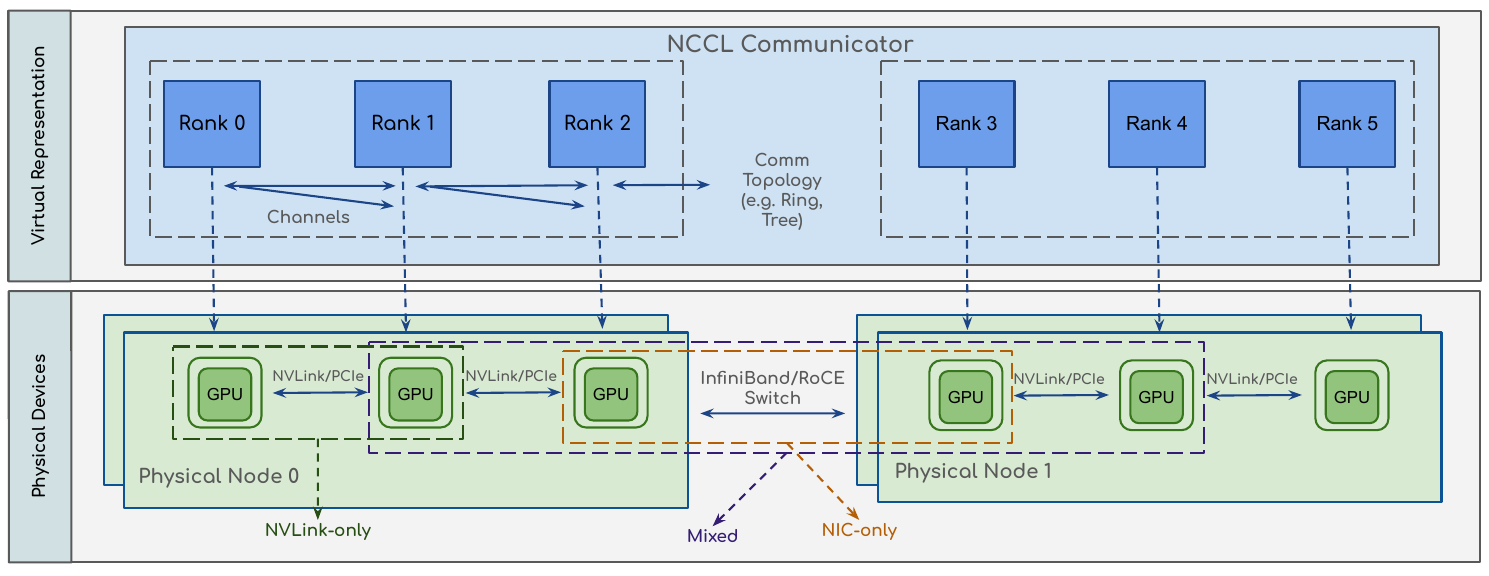}
   \vspace{-3mm}
  \caption{Virtual and physical view of NCCL collective communication. The \emph{virtual representation} (top) shows a communicator that groups ranks across nodes; ranks exchange data along NCCL channels using a chosen logical topology (e.g., ring or tree). The \emph{physical devices} (bottom) show how these ranks are bound to GPUs on different nodes, with intra-node traffic carried over NVLink/PCIe and inter-node traffic over InfiniBand/RoCE through the switch fabric.}
  \label{fig:nccl-comm-logic}
  \vspace{-4mm}
\end{figure*}

Our contributions are as follows:
\begin{itemize}[noitemsep, leftmargin=6mm]
  \item We introduce NIXT, a NCCL Inspector Exporter Tool that improves the observability of collective communication during large-scale ML deployment.  Specifically, NIXT tackles the problem of high volume and dimension of raw NCCL Inspector logs, which limits existing observability approaches.
  \item To support our analysis, NIXT introduces a taxonomy of NCCL Inspector data fields by metric role (identifier, configuration, workload, measurement, counter) and derived properties (e.g., interconnect class) that make collectives interpretable. From the taxonomy we systematically develop summary analysis and correlation analysis (temporal, spatial, resource, other) that fully exploit the NCCL Inspector data.
  \item We perform multiple case studies of NIXT on NCCL Inspector data from production-level LLM pretraining experiments of Nemotron-4, including up to 2,048 GPUs with the 340B model and various GPU configurations with the 15B model. From these case studies, we reveal the performance and variational characteristics of collectives and demonstrate how we can identify targets of potential optimization. We additionally demonstrate case studies for attributing collective performance variation and for localizing and attributing straggler anomalies.
\end{itemize}


%% file: 2.Background.tex
\section{Background}
\label{sec:background_nccl}


The Nvidia Collective Communication Library (NCCL)~\cite{NCCL} implements GPU collectives (AllReduce, AllGather, ReduceScatter, Broadcast) and point-to-point primitives over modern NVIDIA interconnects, mapping operations onto NVLink, PCIe, and RDMA fabrics (InfiniBand, RoCE, EFA). Figure~\ref{fig:nccl-comm-logic} summarizes the organization of NCCL collectives. Ranks (each associated with a single GPU device) are first grouped into \emph{communicators}, which represent the GPUs participating in a collective. Each collective is then parallelized across multiple \emph{channels} (Rank 0-2 and 3-5 are distinct channels). When carrying out a collective, NCCL utilizes these channels by selecting a collective algorithm and logical topology (ring, tree, etc.) based on collective type, message size, and the underlying interconnect. 

NCCL Inspector~\cite{nvidia_nccl_inspector} is a profiler plugin built on the NCCL profiler interface, that transparently samples ongoing NCCL kernels and emits a structured record for every collective operation. To support real-time monitoring of collective statistics, NCCL Inspector also provides a Prometheus interface and Grafana-based live dashboard~\cite{nvidia_inspector_observability_2025,nvidia_inspector_prometheus_2025}. Furthermore, a basic example exporter is provided with NCCL Inspector \cite{nvidia_nccl_inspector} that only provides parsing of log data for statistical summaries and visualization.   
Our work further builds off NCCL Inspector with an advanced Exporter Tool that performs advanced analytics to enable greater observability into collective behaviors and enable actionable insights, such as root cause analysis. 

\paragraph{NCCL Inspector record schema.}
Table~\ref{tab:nccl_inspector_reorganized} lists the fields in each NCCL Inspector record, categorized into a \emph{header} (communicator and rank identifiers), \emph{environment} (timing source, hostname, build), or \emph{collective performance} classification (\texttt{coll}, \texttt{coll\_msg\_size\_bytes}, \texttt{coll\_exec\_time\_us}, \texttt{coll\_algobw\_gbs}, \texttt{coll\_busbw\_gbs})~\cite{NCCL-perf}.  We will use these statistics for the analysis framework in NIXT. We additionally derive another metric \texttt{coll\_type} to identify the type of physical interconnect that is traversed, which we detail later in Section~\ref{sec:framework}.

%% file: 3.Taxonomy.tex

\section{NIXT: NCCL Inspector Exporter Tool}
\label{sec:framework}


\begin{figure*}[!t]
  \centering
  \includegraphics[width=0.75\linewidth]{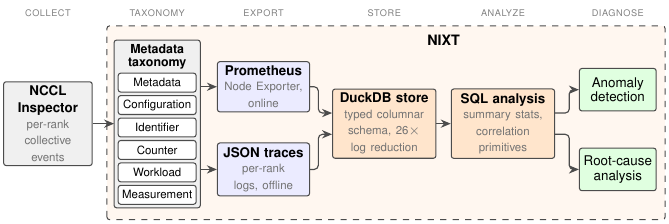}
  \caption{The NIXT NCCL Inspector Exporter Tool workflow}
  \label{fig:Exporter_workflow}
\end{figure*}

In this section, we introduce NIXT, a \underline{N}CCL \underline{I}nspector E\underline{x}porter \underline{T}ool to improve the observability of collective communication during large-scale ML deployment. 
Specifically, NIXT tackles the problem of high volume and dimension of raw NCCL Inspector output. Figure~\ref{fig:Exporter_workflow} overviews the design of NIXT. 

\begin{figure*}[!t]
\centering
\begin{subfigure}[t]{0.193\linewidth}
\includegraphics[width=\linewidth]{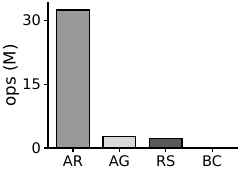}\par\vspace{2pt}
\begin{minipage}[t][6em][t]{\linewidth}
\begin{lstlisting}[style=sqlpaper]
SELECT coll,
  COUNT(*) AS ops
FROM logs
WHERE run='<RUN>'
GROUP BY coll;
\end{lstlisting}
\end{minipage}
\caption{Summary -- Histogram}
\label{fig:analysis-summary}
\end{subfigure}\hfill
\begin{subfigure}[t]{0.193\linewidth}
\includegraphics[width=\linewidth]{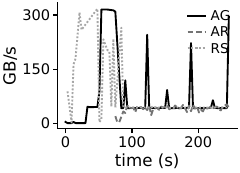}\par\vspace{2pt}
\begin{minipage}[t][6em][t]{\linewidth}
\begin{lstlisting}[style=sqlpaper]
SELECT
  dump_timestamp_us,
  coll, coll_busbw_gbs
FROM logs
WHERE run='<RUN>'
  AND coll_msg_size_bytes
      >= 1000000;
\end{lstlisting}
\end{minipage}
\caption{Temporal correlation}
\label{fig:analysis-temporal}
\end{subfigure}\hfill
\begin{subfigure}[t]{0.193\linewidth}
\includegraphics[width=\linewidth]{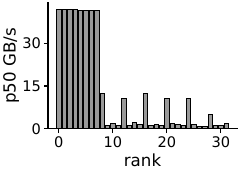}\par\vspace{2pt}
\begin{minipage}[t][6em][t]{\linewidth}
\begin{lstlisting}[style=sqlpaper]
SELECT rank,
  quantile_cont(
    coll_busbw_gbs,
    0.5) AS p50_bw
FROM logs
WHERE run='<RUN>'
  AND coll='AllGather';
\end{lstlisting}
\end{minipage}
\caption{Spatial correlation}
\label{fig:analysis-spatial}
\end{subfigure}\hfill
\begin{subfigure}[t]{0.193\linewidth}
\includegraphics[width=\linewidth]{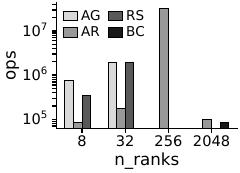}\par\vspace{2pt}
\begin{minipage}[t][6em][t]{\linewidth}
\begin{lstlisting}[style=sqlpaper]
SELECT n_ranks, coll,
  COUNT(*) AS ops
FROM logs
WHERE run='<RUN>'
GROUP BY n_ranks,
         coll;
\end{lstlisting}
\end{minipage}
\caption{Resource correlation}
\label{fig:analysis-resource}
\end{subfigure}\hfill
\begin{subfigure}[t]{0.193\linewidth}
\includegraphics[width=\linewidth]{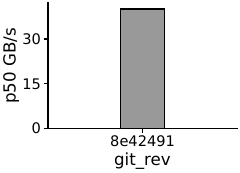}\par\vspace{2pt}
\begin{minipage}[t][6em][t]{\linewidth}
\begin{lstlisting}[style=sqlpaper]
SELECT git_rev,
  quantile_cont(
    coll_busbw_gbs,
    0.5) AS p50_bw
FROM logs
WHERE
  coll='AllGather';
\end{lstlisting}
\end{minipage}
\caption{Other correlation}
\label{fig:analysis-other}
\end{subfigure}

\caption{Analysis primitives implemented by the NCCL Inspector Exporter as SQL queries over the Parquet-backed log store; the \texttt{<RUN>} placeholder is substituted per run. Above each query we show an example rendered output computed over the Nemotron-4 340B @ 2{,}048 GPU log store (the temporal panel is downsampled to the median of each time bin for display).
(a)~\textbf{Summary -- Histogram} on \texttt{coll}: distribution of collective types used in the run.
(b)~\textbf{Temporal correlation} of \texttt{dump\_timestamp\_us} vs \texttt{coll\_busbw\_gbs}: time-varying bandwidth during the workload.
(c)~\textbf{Spatial correlation} of \texttt{rank} vs \texttt{coll\_busbw\_gbs}: median bandwidth achieved by each GPU.
(d)~\textbf{Resource correlation} of \texttt{n\_ranks} vs histogram of \texttt{coll}: how the collective-type distribution shifts with GPU count.
(e)~\textbf{Other correlation} of \texttt{git\_rev} vs \texttt{coll\_busbw\_gbs}: impact of code revision on communication performance.}
\label{tab:analysis}
\end{figure*}

From the raw NCCL Inspector outputs, NIXT implements a unified relational database backend optimized for high-throughput stream processing across both offline JSON traces or live Prometheus network endpoints (provided through NCCL Inspector's Node Exporter feature). The framework translates uncompressed, unstructured logs into a database interface with DuckDB~\cite{duckdb2019} to enable an SQL interface for various summary analysis and correlation analysis, providing observability and facilitating root cause analysis.
This ingestion schema accelerates multi-dimensional SQL querying, allowing complex performance analytics to be processed at scale. Analytical workloads are translated directly into low-overhead relational queries to generate targeted analysis (Figure~\ref{tab:analysis}).

\textbf{Design and implementation.} Beyond ingesting NCCL Inspector JSON into a relational store, NIXT contributes three layers that turn raw profiler output into actionable observability. First, a \emph{taxonomy} of NCCL Inspector fields by metric type (Table~\ref{tab:nccl_inspector_reorganized}, Section~\ref{subsec:taxonomy}) gives the otherwise flat log schema an analysis-oriented structure that makes collectives interpretable. Second, an \emph{analysis-primitive layer} turns the taxonomy into reusable SQL patterns for summary statistics and for temporal, spatial, resource, and other correlation analysis (Figure~\ref{tab:analysis}); each case-study analysis in Sections~\ref{sec:case-comm}--\ref{sec:straggler-case} is an instantiation of these primitives. Third, a \emph{typed columnar schema} provides log reduction: ingest is a single pass that maps every JSON field to a typed column, and DuckDB's columnar compression substantially reduces the on-disk footprint relative to the raw uncompressed JSON.






\begin{table}[b!]
\centering
\footnotesize
\begin{tabular}{|p{0.32\columnwidth}|p{0.35\columnwidth}|p{0.17\columnwidth}|}
\hline
\rowcolor[HTML]{EFEFEF}
\textbf{Metric Name} & \textbf{Definition} & \textbf{Metric Type} \\ \hline
\multicolumn{3}{|l|}{\cellcolor[HTML]{EFEFEF}\textbf{Header}} \\ \hline
id & Unique identifier for the collective operation & Identifier \\ \hline
rank & Logical rank (GPU ID) within the communicator & Identifier \\ \hline
n\_ranks & Total number of GPUs in this collective & Configuration \\ \hline
nnodes & Number of physical compute nodes involved & Configuration \\ \hline
coll\_type & Derived topology type (e.g., nvlink-only) & Configuration \\ \hline
\rowcolor[HTML]{EFEFEF}
\multicolumn{3}{|l|}{\cellcolor[HTML]{EFEFEF}\textbf{Environment}} \\ \hline
inspector\_output\_\newline format\_version & Version of the output schema & Metadata \\ \hline
git\_rev & Git revision of the build & Metadata \\ \hline
rec\_mechanism & The recording method & Metadata \\ \hline
dump\_timestamp\_us & Microseconds epoch timestamp & Counter \\ \hline
hostname & Name of the host machine & Identifier \\ \hline
pid & OS process ID & Identifier \\ \hline
coll\_timing\_source & Source of timing (e.g., kernel\_gpu) & Metadata \\ \hline
\rowcolor[HTML]{EFEFEF}
\multicolumn{3}{|l|}{\cellcolor[HTML]{EFEFEF}\textbf{Collective Performance (coll\_perf)}} \\ \hline
coll & Type of collective (e.g., AllReduce) & Workload \\ \hline
coll\_sn & Sequence number of the call & Counter \\ \hline
coll\_msg\_size\_bytes & Message size in bytes & Workload \\ \hline
coll\_exec\_time\_us & Total execution time ($\mu$s) & Measurement \\ \hline
coll\_algobw\_gbs & Algorithmic bandwidth (GB/s) & Measurement \\ \hline
coll\_busbw\_gbs & Hardware bus bandwidth (GB/s) & Measurement \\ \hline
\end{tabular}
\caption{NCCL Inspector data fields. 
}
\label{tab:nccl_inspector_reorganized}
\end{table}

\subsection{Taxonomy of NCCL Inspector Data}
\label{subsec:taxonomy}
To facilitate our analysis for observability, we introduce a taxonomy of the NCCL Inspector metric in Table~\ref{tab:nccl_inspector_reorganized} with a \textit{metric type} as following: 

\begin{itemize}[noitemsep]
\item \textit{Metadata}: Software specification that is consistent within the current NCCL Inspector.  
\item \textit{Configuration}: Fields that describe how a run is set up, including communicator shape (for example, number of ranks or nodes). Configurations correspond to \textit{resource} properties of the collective.
\item \textit{Identifier}: Static fields that uniquely label entities such as ranks, processes, or hosts (for example, rank ID, hostname, or PID). Identifiers correspond to \textit{spatial} properties of the collective. 
\item \textit{Counter}: Monotonic or event-like values that record counts or timestamps. Counters correspond to \textit{temporal} properties of the collective. 
\item \textit{Workload}: Fields that describe workload communication properties, such as collective type and message size. 
\item \textit{Measurement}: Quantitative performance metrics directly observed or derived from execution, i.e. latency and bandwidth. 
\end{itemize}

We will use these taxonomies in our analysis to identify correlation and root cause analysis. Additionally, we derive an interconnect classification metric. NCCL Inspector does not expose packet-level transport details (i.e. the type of link that is traversed), however, it can be derived. Using communicator placement (\texttt{n\_ranks}, \texttt{nnodes}) we can infer whether collectives traverse intra-node links, inter-node fabric, or a mixture. 

We derive a class \texttt{coll\_type} for each collective, with four values: \emph{single-rank} (\texttt{n\_ranks}=1), \emph{nvlink-only} (\texttt{nnodes}=1; all ranks within one node), \emph{NIC-only} (\texttt{n\_ranks}=\texttt{nnodes}; one rank per node, dominated by the host channel adapter), and \emph{mixed} (multi-node with multiple ranks per node). These coarse classes are used throughout our analysis to connect observed bandwidth to physical interconnect type.


\subsection{Analysis Methodology}
\label{subsec:methodology}

Broadly, we classify our
analysis insights into two categories: 
\textit{summary statistics} and multi-variate \textit{correlation analysis}. We present examples of our analysis, and associated SQL queries in Figure~\ref{tab:analysis}.

\subsubsection{Summary Statistics} 
~\\Summary statistics provides us with traditional statistical measurements that summarizes a single metric. Common summary statistics for numerical metrics include central tendencies (mean, median), spread (standard deviation, spread, quartiles) and distributions (histograms). For categorical metrics (such as collective type), summary statistics can also provide us with distributions (for example, how often each collective type are used). For example, Figure~\ref{fig:analysis-summary} obtains a histogram of collective types during a workload.

\subsubsection{Correlation Analysis} 
~\\While summary statistics applies analysis to a single metric, deeper insights can be derived from correlation analysis where multiple metrics are analyzed together. We categorize correlation analysis into temporal correlation, spatial correlation, resource correlation, and other correlation.

\textbf{Temporal correlation analysis} compares a timing-based metric (Counter metrics, such as timestamp and sequence number) with another metric of interest. For example, timestamp can be correlated with bandwidth metrics to identify time-varying bandwidth behaviors of the workload. Furthermore, timestamp can be correlated with collective type to observe patterns of collective calls which can potentially help identify and fingerprint phases of workloads. Figure~\ref{fig:analysis-temporal} shows the bandwidth over time for each collective type of message sizes larger than 1MB. 

\textbf{Spatial correlation analysis} compares a spatial-related metric (such as rank, hostname, pid, or other Identifier metric.) with a metric of interest. This allows us to identify spatial-related relationships in the data. For example, when looking at the correlation between rank (GPU ID) and bandwidth, we can potentially identify anomalies or bottlenecks that impact specific GPUs. Figure~\ref{fig:analysis-spatial} shows a correlation analysis of the All Gather bandwidth achieved by each GPU. 

\textbf{Resource correlation analysis} is a use case based on configuration-type metrics (such as number of GPUs or number of nodes). For example, number of nodes can be correlated against collective performance metrics to gain insight into how the workload's collective communication performance scale as workload scales. As another example, when correlated with distribution of collective types, we can see how the proportion of various collective type change as workload scales. Figure~\ref{fig:analysis-resource} shows how collective type distribution change
with number of GPUs used. 

Resource correlation analysis is especially important because when resources scale, there can be many hidden effects on low-level collective properties that NCCL Inspector can provide unique observability in to. 
For example, in smaller configurations (e.g., Nemotron-4 15B on 16 GPUs), only a few communicator shapes and message sizes are exercised, corresponding to simple tensor-parallel and data-parallel groups. 
As we increase the number of GPUs and model size (e.g., Nemotron-4 340B), additional data‑parallel replicas are added to the model that increase the communicator world size and more complex collectives. The additional pipeline and virtual pipeline stages in larger models introduce more communicator patterns, and the distinction between intra-node and inter-node communication becomes pronounced.

\textbf{Other correlation analysis} that don't fall within spatial, temporal, or resource aspects are equally important in deriving insights into workloads. For example, correlating Git revision of build and bandwidth can give us insight into software performance regressions. Also, correlating any Measurement-based metric with Identifier-based metric can potentially be useful in anomaly detection. Figure~\ref{fig:analysis-other} shows how the code version impact  communication performance.

\subsection{Anomaly and Root Cause Analysis}
Using the aforementioned summary and correlation analysis, NIXT is also able to perform anomaly and root cause analysis to aid in collective communication debugging of large-scale ML deployment. As we will demonstrate in Section~\ref{sec:nccl-tests-baseline} and~\ref{sec:straggler-case}, we will describe how combinations of the aforementioned correlation analysis can be used to aid in attribution of collective variation and in localization and attribution of GPU straggler anomalies. These case studies will demonstrate how NIXT's extended observability of NCCL statistics enables novel use cases of advanced analysis. 


\subsection{Overhead}
\label{subsec:nixt-overhead}
NIXT is deliberately designed to stay off the training critical path. The only runtime cost is enabling NCCL Inspector profiling itself, which we measured at under 2\% end-to-end overhead for real training workloads across scales from 16 to 2{,}048 GPUs, with validation across all major collective operations. NIXT then ingests the resulting JSON records into DuckDB post-hoc, so training throughput is unaffected by the analysis layer.

For the largest trace in our dataset (Nemotron-4 340B at 2{,}048 GPUs), single-pass ingest completes in under a minute (\textbf{56}~seconds end-to-end, \textbf{24}~seconds for the Parquet-write stage on a 64-core commodity node), and the resulting DuckDB store occupies \textbf{0.67}~GB versus \textbf{17.7}~GB of raw uncompressed JSON (a \textbf{26$\times$} reduction). The analysis primitives of Figure~\ref{tab:analysis} execute in \textbf{under 0.3 seconds each} on this store on a single node. A single-node backend is sufficient at 2{,}048-GPU trace volume; for multi-day runs, the store can be partitioned by run and time window, with cold partitions attached as read-only Parquet.


\section{Experiment Methodology}
\label{subsec:exp_setup}

\textbf{Hardware and Cluster Configuration. }
\label{sec:exp_setup:hardware}
All experiments were conducted on a multi-node NVIDIA H100 GPU cluster.
Each node is an 8-GPU NVIDIA HGX H100 platform with high-bandwidth intra-node GPU interconnect (NVLink/NVSwitch), and nodes are connected by a high-speed network via host channel adapters (NICs) (e.g., InfiniBand). Our runs scale up to 256 nodes (2{,}048 GPUs).

\textbf{Model Workload. }
\label{sec:exp_setup:software}
For our case study, we run a Nemotron-4 pretraining workloads implemented in NVIDIA NeMo (Megatron-core backend), focusing on two representative model size: Nemotron-4 15B and Nemotron-4 340B. We ran Nemotron-4 15B with 16, 32, 64, 128, and 2048 GPUs with data parallelism and tensor parallelism (TP=2). We ran Nemotron-4 340B with 2048 GPUs with data parallelism and tensor parallelism (TP=8), pipeline parallelism (PP=8), and virtual pipeline parallelism (VP=12). In the model training, we use NCCL 2.28 and the default NCCL Inspector shipped with it.

For the 15B model, the model parallel requirement is not large so two H100 GPU is sufficient to hold the whole model. Therefore, GPU scaling mainly changes the size of data parallelism.
For the 340B model, it requires more GPU memory to host the model, thus tensor parallelism is used for intra-node and pipeline parallelism is used inter-node. We also use virtual pipeline parallel, which lets each physical pipeline stage behave as multiple logical stages and improves utilization by decreasing idle bubbles.


%% file: 4.CaseStudy.tex
\section{Case Study: Observability into Communication Behavior in Large-Scale LLM Training}
\label{sec:case-comm}
This section demonstrates how NIXT is able to analyze real LLM training workloads on a production GPU cluster. We use these profiling traces to characterize collective communication behavior and to derive insights.

\subsection{Summary Statistics Analysis}
\label{sec:results:summary}
We first demonstrate an example of basic summary statistics analysis on Nemotron-4 340B at 2{,}048 GPUs. This configuration combines data parallelism with tensor parallelism (TP=8), pipeline parallelism (PP=8), and virtual pipeline parallelism (VP=12), and therefore exercises the broadest mix of communicators and link types in our dataset. 

\begin{figure}[t!]
  \centering
  \includegraphics[width=0.75\linewidth]{\detokenize{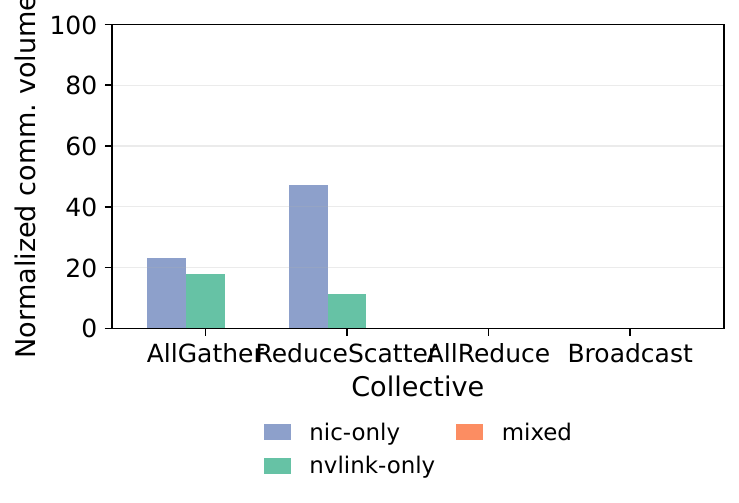}}
  \vspace{-3mm}
  \caption{Nemotron-4 340B @ 2{,}048 GPUs: total communicated bytes by collective type and communicator class.}
  \label{fig:340b_bytes_mix}
\end{figure}

Figure~\ref{fig:340b_bytes_mix} shows the volume of communication by each
collective type, broken down by communicator class, normalized by the total
communication volume observed.

\begin{tcolorbox}[colback=lightgreen!20!white,colframe=lightgreen!50!black,
                  left=2pt,right=2pt,top=2pt,bottom=2pt]
\textbf{Observation:} ReduceScatter and AllGather account for nearly all communication volume (58.5\% and 41.2\%, respectively), while AllReduce and Broadcast contribute less than 0.4\%. Across communicator classes, 70.4\% of bytes traverse NIC-only paths, 29.6\% remain nvlink-only, and mixed paths carry a negligible 0.05\%.
\end{tcolorbox}
While summary statistics can give us an overview of collective characteristics, we now demonstrate how NIXT is able to provide deeper observability through multi-variate correlation analysis.



\subsection{Resource Correlation Analysis}
We first demonstrate examples of resource correlations, which are derived from correlation analysis that utilize \textit{Configuration} metric type, which corresponds to the hardware resources used, such as communicator size (\texttt{n\_ranks}) and number of nodes (\texttt{nnodes}).


\subsubsection{Observability into communicator topology and message size}
\begin{figure}[t!]
  \centering
  \begin{subfigure}[t]{0.99\linewidth}
    \centering
    \includegraphics[width=\linewidth]{\detokenize{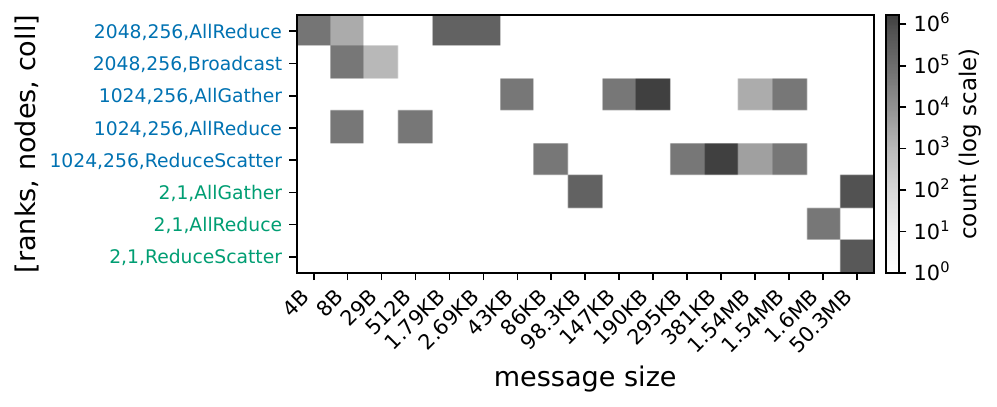}}
    \vspace{-7mm}
    \caption{Nemotron-4 15B @ 2{,}048 GPUs}
    \label{fig:config-15b-2048}
  \end{subfigure}

  \begin{subfigure}[t]{0.99\linewidth}
    \centering
    \includegraphics[width=\linewidth]{\detokenize{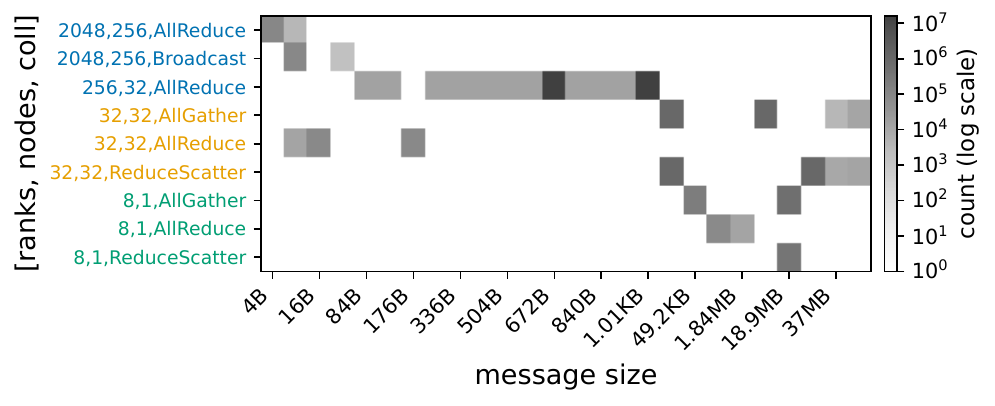}}
    \vspace{-7mm}
    \caption{Nemotron-4 340B @ 2{,}048 GPUs}
    \label{fig:config-340b-2048}
  \end{subfigure}

  \caption{Resource correlation analysis of communicator topology vs message size. \textcolor{ForestGreen}{Green} indicates nvlink-only, \textcolor{orange}{orange} is NIC-only, and \textcolor{blue}{blue} is mixed communication links. }
  \label{fig:config-heatmaps}
\end{figure}

We define \textit{communicator topology} as the tuple of [\texttt{n\_ranks}, \texttt{nnodes}, \texttt{coll}] and we look at the correlation with message size (\texttt{coll\_msg\_size\_bytes}). 
Figures~\ref{fig:config-15b-2048} and~\ref{fig:config-340b-2048} illustrates a heatmap showing the count for a given message size (x-axis) and communicator topology (y-axis) for a single run of Nemotron-4 15B and 340B, respectively, with 2,048 GPUs. The communicator topology shown are all observed communicator topology for pre-training run. 

\begin{tcolorbox}[colback=lightgreen!20!white,colframe=lightgreen!50!black,
                  left=2pt,right=2pt,top=2pt,bottom=2pt]
\textbf{Observation:} 
Only a small number of communicator topology and message size combinations account for the vast majority of collective calls. These ``hot'' configurations identify potential targets for further optimization.
\end{tcolorbox}

For the 15B model (Figure~\ref{fig:config-15b-2048}), we see only a handful of active configuration rows. Nvlink-only \([2,1,*~\footnote{~\(*\) indicates a wildcard for a given field. }]\) AllGather and ReduceScatter buckets and NIC-only \([1024,256,*]\) AllGather/ReduceScatter buckets dominate, each concentrated at one or two message sizes (e.g., 49–50~MiB intra-node and 13–27~MiB inter-node transfers). Global \([2048,256,*]\) AllReduce and Broadcast calls appear at very small message sizes and with lower counts, reflecting their role as control and coordination collectives rather than data movement.

The 340B model (Figure~\ref{fig:config-340b-2048}) exercises more rows, corresponding to additional tensor-parallel and pipeline-parallel groups, but the pattern remains sparse and structurally similar to 15B. Nvlink-only \([8,1,*]\) collectives implement intra-node tensor-parallel communication; NIC-only \([32,32,*]\) AllGather/ReduceScatter handle the main data-parallel traffic; intermediate \([256,32,\text{AllReduce}]\) configurations are used inside the model-parallel hierarchy; and global \([2048,256,*]\) AllReduce/Broadcast calls again appear with small message sizes. In each band, a small number of message sizes dominate the color scale, indicating that Nemotron-4 repeatedly uses a few fixed buckets per communicator type.


\begin{figure}[!b]
  \centering
  \includegraphics[width=\linewidth]{\detokenize{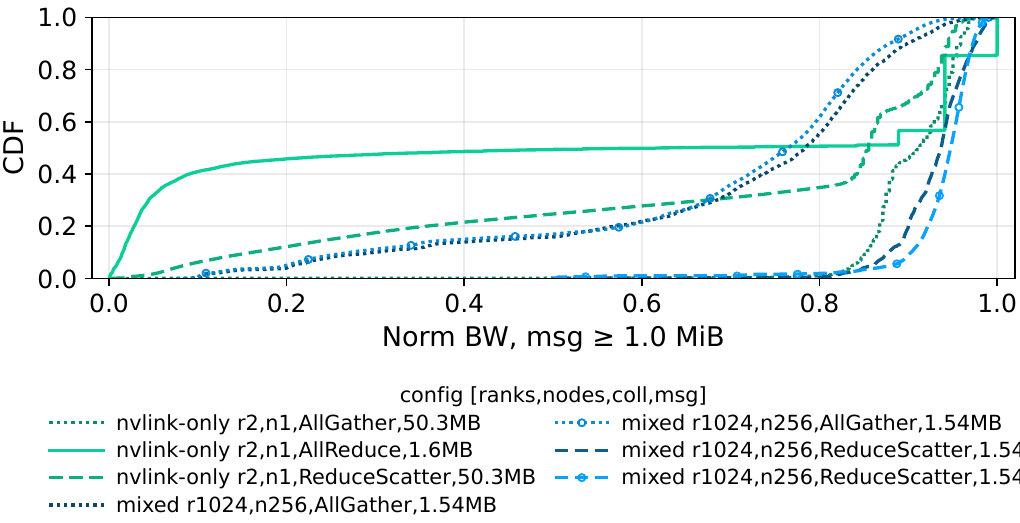}}
  \caption{Bus-bandwidth ECDFs for dominant large-message communicator topology (Nemotron-4 15B @ 2{,}048 GPUs). Each curve is individually normalized to the maximum bandwidth observed for its communicator topology and message size.}
  \label{fig:ecdf-bw-15b_gpus2048}
\end{figure}

\begin{figure}[!t]
  \centering
  \includegraphics[width=\linewidth]{\detokenize{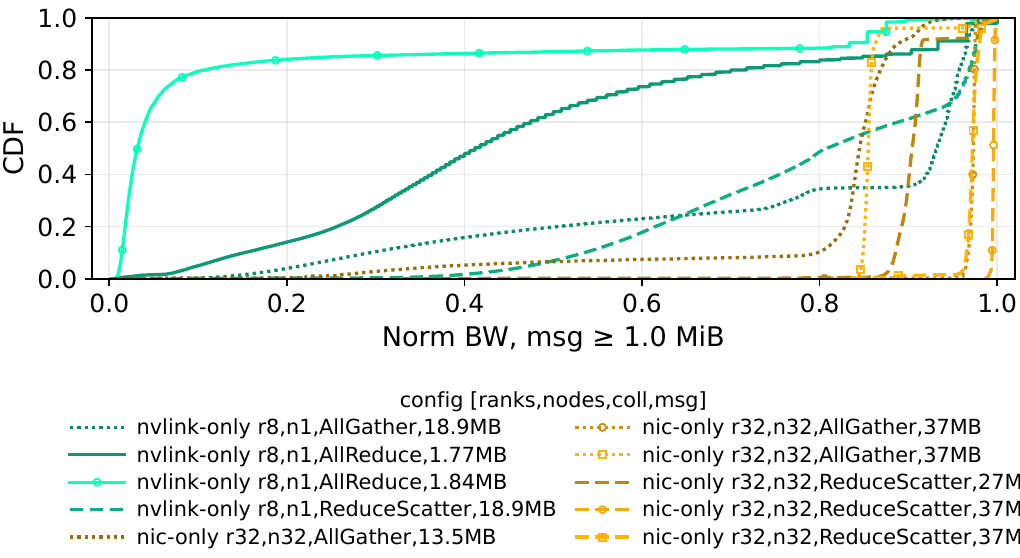}}
  \caption{Bus-bandwidth ECDFs for dominant large-message communicator topology (Nemotron-4 340B @ 2{,}048 GPUs). Each curve is individually normalized to the maximum bandwidth observed for its communicator topology and message size.}
  \label{fig:ecdf-bw-340b_gpus2048}
  \vspace{-3mm}
\end{figure}

\subsubsection{Performance Observability of Common Communicator Topologies}
\label{subsubsec:dominant-config-performance}


We now examine how the dominant communicator topologies from Figure~\ref{fig:config-heatmaps} perform in terms of bus bandwidth and execution time for both Nemotron-4 15B and 340B at 2{,}048 GPUs. 
Figures~\ref{fig:ecdf-bw-15b_gpus2048} and~\ref{fig:ecdf-bw-340b_gpus2048} present empirical CDFs of selected bus bandwidth for large-message collectives in each run. 

\begin{tcolorbox}[colback=lightgreen!20!white,colframe=lightgreen!50!black,
                  left=2pt,right=2pt,top=2pt,bottom=2pt]
\textbf{Observation:} A small set of communicator topology and message size combinations --- the ``hot'' kernels --- carry the bulk of collective traffic, and each exhibits a stable, characteristic bandwidth distribution. Identifying these kernels pinpoints exactly which collectives are worth optimizing.
\end{tcolorbox}

For the Nemotron-4 15B model, nvlink-only \([2,1,\texttt{AllGather}/\texttt{ReduceScatter}]\) communicators with 50MB messages reach high median bandwidths with low variation, while NIC-only \([1024,256,*]\) communicators with 13--27MB achieve lower medians and slightly longer tails. The 340B model exhibits the same qualitative pattern for its dominant nvlink-only \([8,1,*]\) 18.9MB and NIC-only \([32,32,*]\) 13--27MB communicators. Model size influences these distributions indirectly: it determines the parallelism configuration (TP/PP/DP), which in turn determines the communicator topologies and message sizes that are exercised and how often. Once a (topology, message size) kernel is fixed, its bandwidth distribution is largely insensitive to which model produced it.

\begin{figure}[t]
  \centering
  \includegraphics[width=\linewidth]{\detokenize{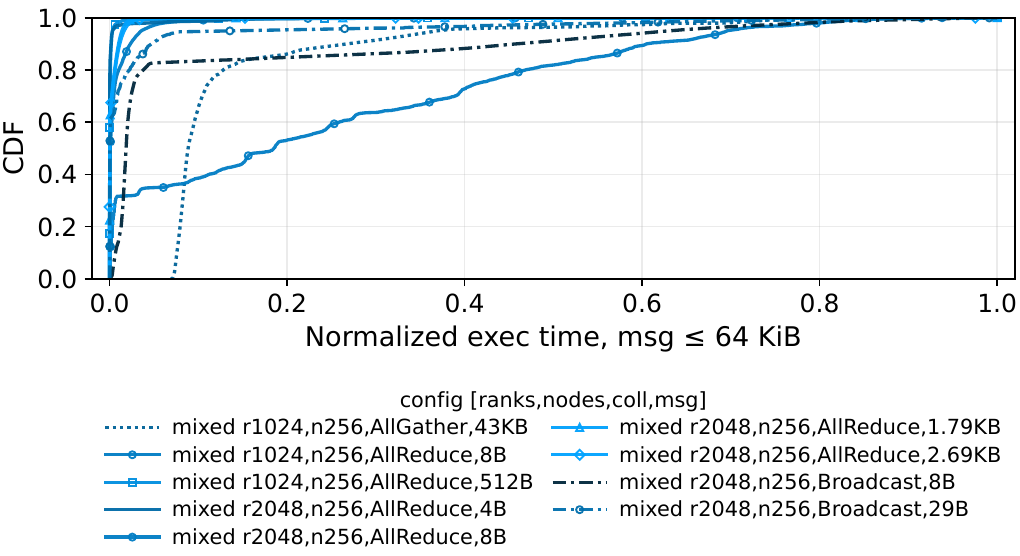}}
  \caption{Execution-time ECDFs for dominant small-message communicator topology (Nemotron-4 15B @ 2{,}048 GPUs).}
  \label{fig:ecdf-latency-15b_gpus2048}
\end{figure}

\begin{figure}[t]
  \centering
  \includegraphics[width=\linewidth]{\detokenize{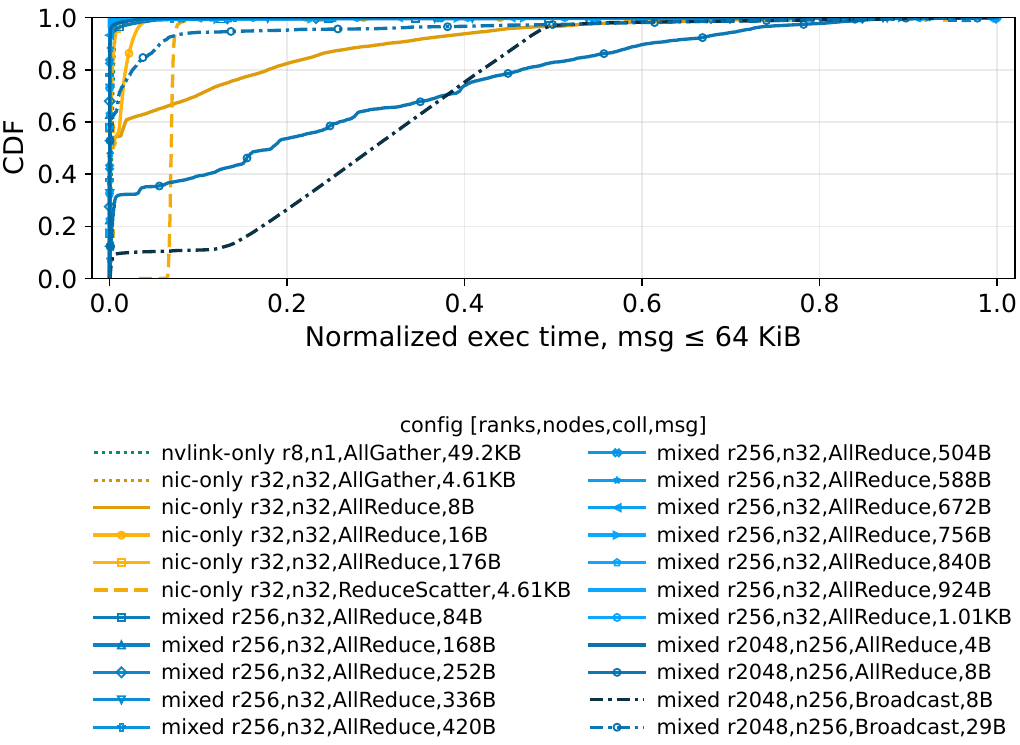}}
  \caption{Execution-time ECDFs for dominant small-message communicator topology (Nemotron-4 340B @ 2{,}048 GPUs).}
  \label{fig:ecdf-latency-340b_gpus2048}
\end{figure}

Across both models and message regimes, the ECDFs reveal clear “sweet spots’’: large-message nvlink-only and NIC-only communicator topologies deliver stable, near-saturating bus bandwidth, while small-message global AllReduce/Broadcast communicators (Figure~\ref{fig:ecdf-latency-15b_gpus2048} and Figure~\ref{fig:ecdf-latency-340b_gpus2048}) provide consistently lower latency. The key takeaway of this analysis is that improving end-to-end scalability comes down to optimizing a small set of configuration-specific kernels, rather than the full combinatorial configuration space --- and NIXT identifies exactly which kernels those are for a given workload.

\subsubsection{Observability into Scaling GPU Count}

\begin{figure}[t]
  \centering
  \begin{subfigure}[t]{0.49\linewidth}
    \centering
    \includegraphics[width=\linewidth]{\detokenize{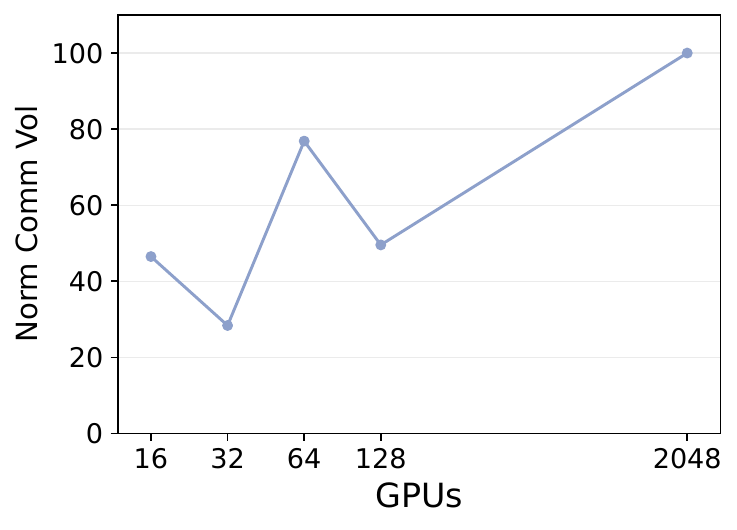}}
    \caption{Total bytes vs GPUs}
    \label{fig:15b-scale-total}
  \end{subfigure}\hfill
  \begin{subfigure}[t]{0.49\linewidth}
    \centering
    \includegraphics[width=\linewidth]{\detokenize{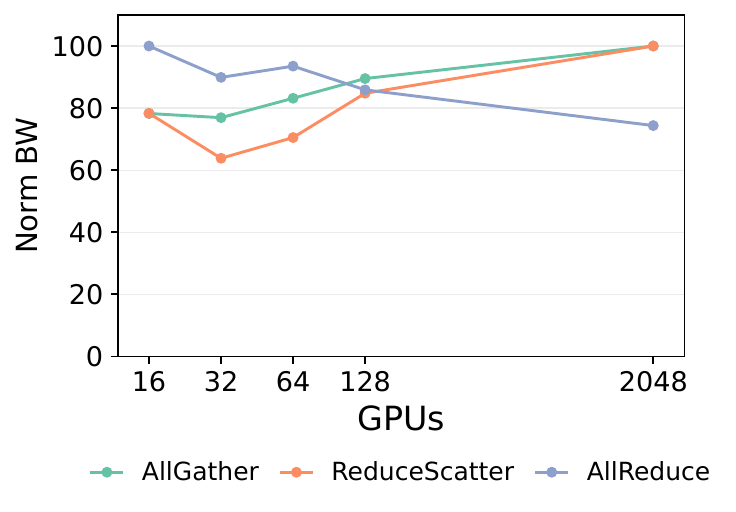}}
    \caption{Bus BW vs GPUs}
    \label{fig:15b-scale-bw}
  \end{subfigure}
  \caption{Nemotron-4 15B scaling with GPU count. In (a), NCCL Inspector samples collectives rather than recording every operation, so the captured volume varies across runs; this sampling accounts for the non-monotonic pattern, while the overall trend grows with GPU count.}
  \label{fig:15b-scale-pair}
  \vspace{-4mm}
\end{figure}

\begin{figure}[t]
  \centering
  \includegraphics[width=0.98\linewidth]{\detokenize{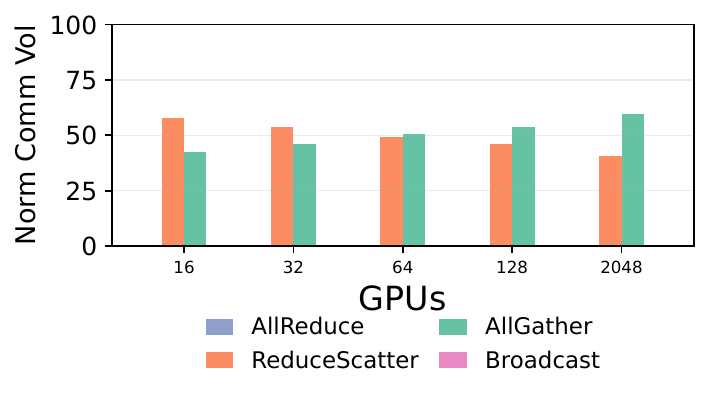}}
  \caption{Nemotron-4 15B: communicated bytes by collective type versus GPU count.}
  \label{fig:15b-scale-by-coll}
  \vspace{-4mm}
\end{figure}

We now study strong-scaling behavior for a fixed model size by running Nemotron-4 15B at 16, 32, 64, 128, and 2{,}048 GPUs with identical tensor-parallel configuration (TP = 2). In this setting, increasing the number of GPUs primarily increases the data-parallel world size and the number of pipeline groups, while each replica of the model structure remains unchanged.

Figure~\ref{fig:15b-scale-total} shows that the total communication volume (in bytes) captured in the training window grows with GPU count. Because NCCL Inspector samples collectives rather than recording every operation, the captured volume varies from run to run, which produces the non-monotonic pattern in the figure; the underlying trend is nonetheless clearly increasing.
This reflects both the additional data-parallel replicas that must synchronize and the deeper hierarchy of communicators that appears at larger scales. Figure~\ref{fig:15b-scale-by-coll} breaks this volume down by collective type. 
While AllGather and ReduceScatter dominate at all scales, their volume change in opposite directions when the model scales with more GPUs. This shows the increased data parallel greatly increase the usage of AllGather collectives for synchronize of the model parameters.

Figure~\ref{fig:15b-scale-bw} shows that the volume-weighted mean bus bandwidth for the dominant AllGather and ReduceScatter collectives is remarkably stable across GPU counts. The median bus bandwidth for large-message communicators changes only slightly between 16 and 2{,}048 GPUs, and the relative occurrence of collective types is preserved.

\begin{tcolorbox}[colback=lightgreen!20!white,colframe=lightgreen!50!black,
                  left=2pt,right=2pt,top=2pt,bottom=2pt]
\textbf{Observation :} For Nemotron-4 15B, increasing GPU count mainly scales the communication portion of AllGather, confirming that data parallelism heavily relies on AllGather collective. 
\end{tcolorbox}

The increased bus bandwidth of AllGather when GPU scales shows that increasing data parallelism can benefit the collective efficiency of training.
This behavior is consistent with the configuration-space view: as we add GPUs, the workload instantiates more instances of the same hot communicator topology (e.g., NIC-only and nvlink-only AllGather/ReduceScatter communicators), rather than shifting to fundamentally slower communicator topology. Consequently, the communication cost of scaling 15B to thousands of GPUs is dominated by the aggregate number and size of these repeated collective calls, not by a deterioration in the throughput of individual collectives.

\subsubsection{Observability with Scaling Model Size}
\begin{figure}[t]
  \centering
  \begin{subfigure}[t]{0.49\linewidth}
    \centering
    \includegraphics[width=\linewidth]{\detokenize{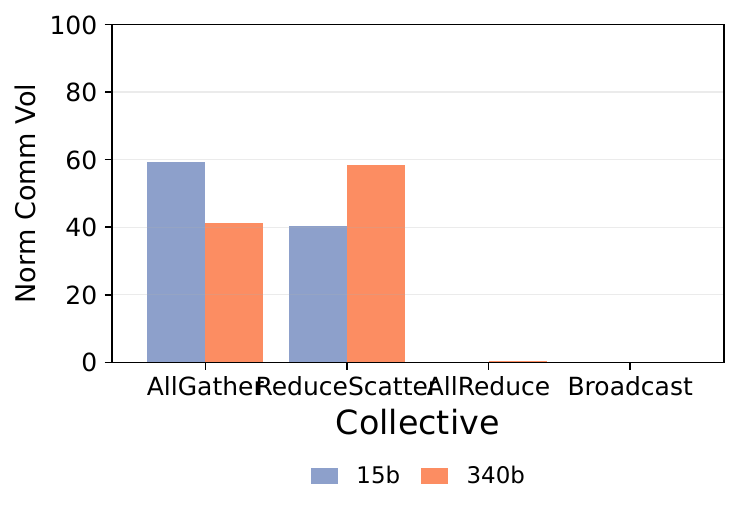}}
    \caption{Communicated bytes by collective (log bytes).}
    \label{fig:compare-modelsize-2048-bytes}
  \end{subfigure}\hfill
  \begin{subfigure}[t]{0.49\linewidth}
    \centering
    \includegraphics[width=\linewidth]{\detokenize{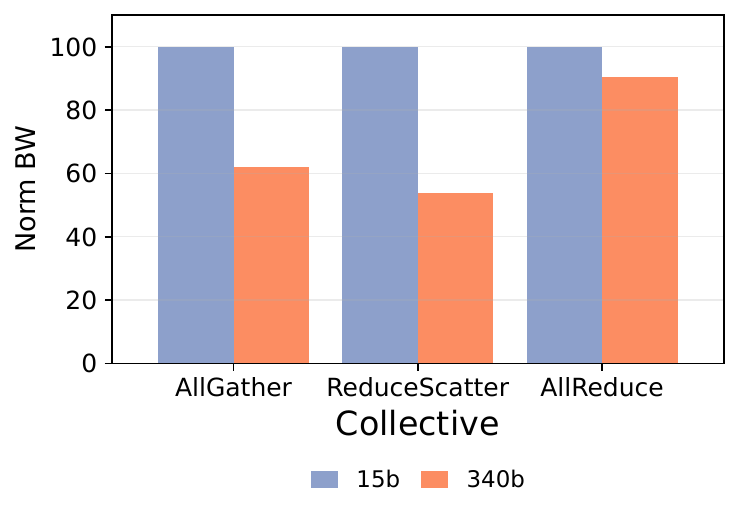}}
    \caption{Volume-weighted mean bus BW ($\ge 1$\,MB messages).}
    \label{fig:compare-modelsize-2048-busbw}
  \end{subfigure}
  \caption{Nemotron-4 at 2{,}048 GPUs: 15B versus 340B comparison of (a) communicated bytes by collective type and (b) volume-weighted mean bus bandwidth by collective.}
  \label{fig:compare-modelsize-2048-pair}
  \vspace{-2mm}
\end{figure}

We now compare two model sizes (Nemotron-4 15B and 340B) at 2{,}048 GPUs. Figure~\ref{fig:compare-modelsize-2048-pair}(a) shows communicated bytes per collective type and Figure~\ref{fig:compare-modelsize-2048-pair}(b) shows volume-weighted mean bus bandwidth per collective (messages $\ge 1$\,MB). 

\begin{tcolorbox}[colback=lightgreen!20!white,colframe=lightgreen!50!black,
                  left=2pt,right=2pt,top=2pt,bottom=2pt]
\textbf{Observation:} Increasing model size at fixed GPU count shifts the communication \emph{mix} (340B exercises a richer set of collective/communicator combinations, with NIC-only carrying most bytes). Thus, we see a bandwidth drop because more collective traffic goes through the inter-node network rather than intra-node NVLink.

\end{tcolorbox}

\begin{figure}[!b]
  \centering
  \vspace{-3mm}
  \includegraphics[width=\linewidth]{\detokenize{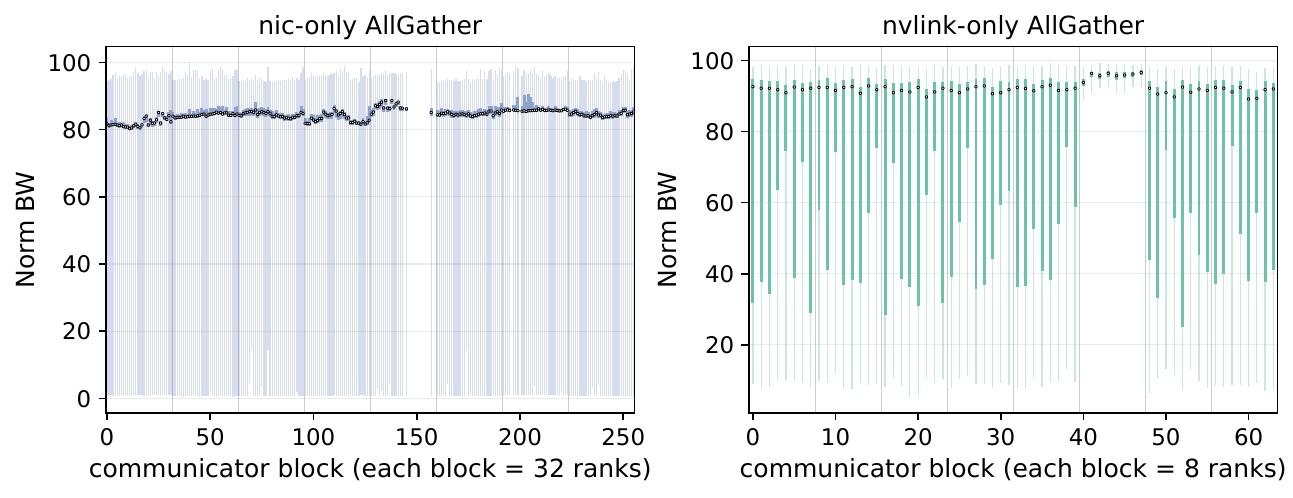}}
  \caption{Bus bandwidth at every (communicator, rank) cell (Nemotron-4 340B @ 2{,}048 GPUs). Each communicator's ranks are laid out as a contiguous block; thin gray dividers separate communicators. Whiskers span min--max, the thick segment spans p25--p75.}
  \label{fig:spatial-percommrank-rangebar}
\end{figure}

\subsection{Spatial Correlation Analysis}

Here, we demonstrate spatial correlation analysis, where the identifier dimension addresses \emph{where} certain patterns occur in the cluster by correlating measurements with logical ranks and hostnames. This spatial view is essential for spotting stragglers and localizing anomalies that are invisible in configuration-averaged statistics.

\subsubsection{Observability into Identifier-dependent Performance Variation}

Figure~\ref{fig:spatial-percommrank-rangebar} shows the normalized bandwidth with range bars indicating 25th/75th percentile (thicker bars) and min/max (thin whiskers) of the two most common communicator topologies, nic-only AllGather 32 rank/32 node at \textasciitilde13MB (inter-node data-parallel) and nvlink-only AllGather 8 rank/node at \textasciitilde18MB (intra-node tensor-parallel). The X-axis is indexed by communicator ID, blocked into 8 or 32 ranks for readability.

\begin{tcolorbox}[colback=lightgreen!20!white,colframe=lightgreen!50!black,
                  left=2pt,right=2pt,top=2pt,bottom=2pt]
\textbf{Observation:} Bandwidth variation exhibits identifier-centric behavior. Certain communicators (between communicator block IDs 40-50) achieve higher bandwidth and less variation compared to all other devices.
\end{tcolorbox}

This demonstrates that identifier information can improve the observability into spatial correlations. Here we identify that certain devices achieve higher throughput with less performance variability, despite running the same collective (AllGather) over the same physical links (NVLink) of the same message size (\textasciitilde18MB), indicating device-level performance variation. NIXT localizes \emph{which} devices behave differently; attributing the difference to a specific cause --- thermal state, hardware revision, or software-level effects such as load imbalance --- requires correlating with device telemetry (e.g., DCGM and driver-level metrics), which we discuss as future work. Similarly, this analysis can also provide insight into negative-impact anomaly detection.


\begin{figure}[!t]
  \centering
  \includegraphics[width=\linewidth]{\detokenize{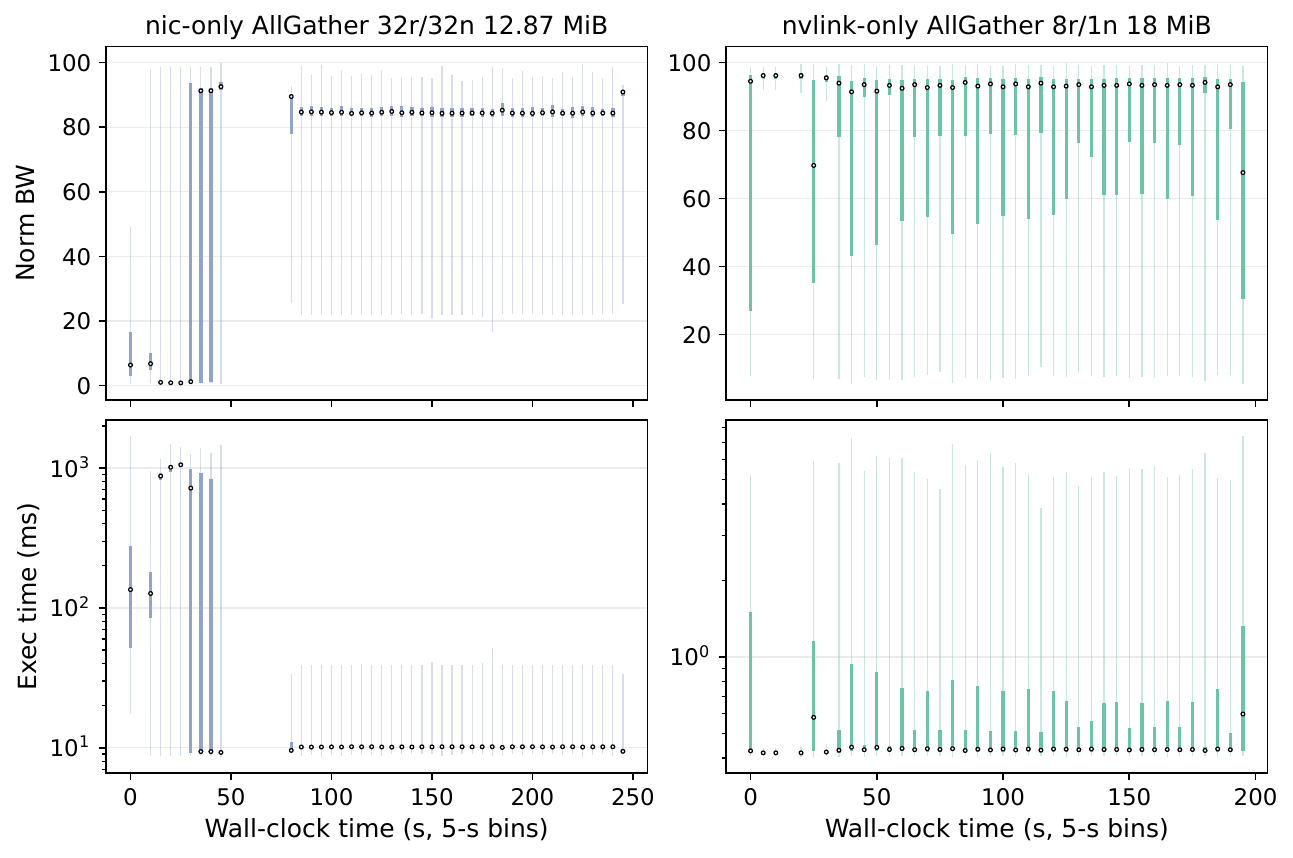}}
  \caption{Bus bandwidth and exec time vs wall-clock time, binned at 5\,s (Nemotron-4 340B @ 2{,}048 GPUs). 
  }
  \label{fig:temporal-time-rangebar}
\end{figure}

\subsection{Temporal Correlation Analysis}

Here, we demonstrate temporal correlation analysis, where the counter dimension addresses \emph{when} certain patterns occur in the cluster by correlating measurements with sequence number and timestamps. 
Temporal correlation analysis can benefit the spotting of new occurrence of error or performance degradation.


In Figure~\ref{fig:temporal-time-rangebar}, two collectives are selected and displayed by timestamp and sequence number (x-axis).
The left side are NIC-only AllGather on 32 ranks, 32 nodes, with message size of 12.87MB, and the right side are NVLink-only AllGather on 8 ranks, 1 node, with message size of 18MB. 
Collectives are grouped every 50 operations for all the communicators for readability.

On the left of Figure~\ref{fig:temporal-time-rangebar}, we see the bandwidth of the NIC-only AllGather in the first 0-25s of this collective go through a low value, low variance phase, then during 25-50s it has higher range of values but still high variance, only after around 70s it stabilizes to a fixed high bandwidth with low variance range. This might be connected to the network contention in the initialize phase.
On the right of Figure~\ref{fig:temporal-time-rangebar}, we don't see this unstable initial phase for the NVLink-only AllGather within one node.

\begin{figure}[b!]
  \centering
  \includegraphics[width=\linewidth]{\detokenize{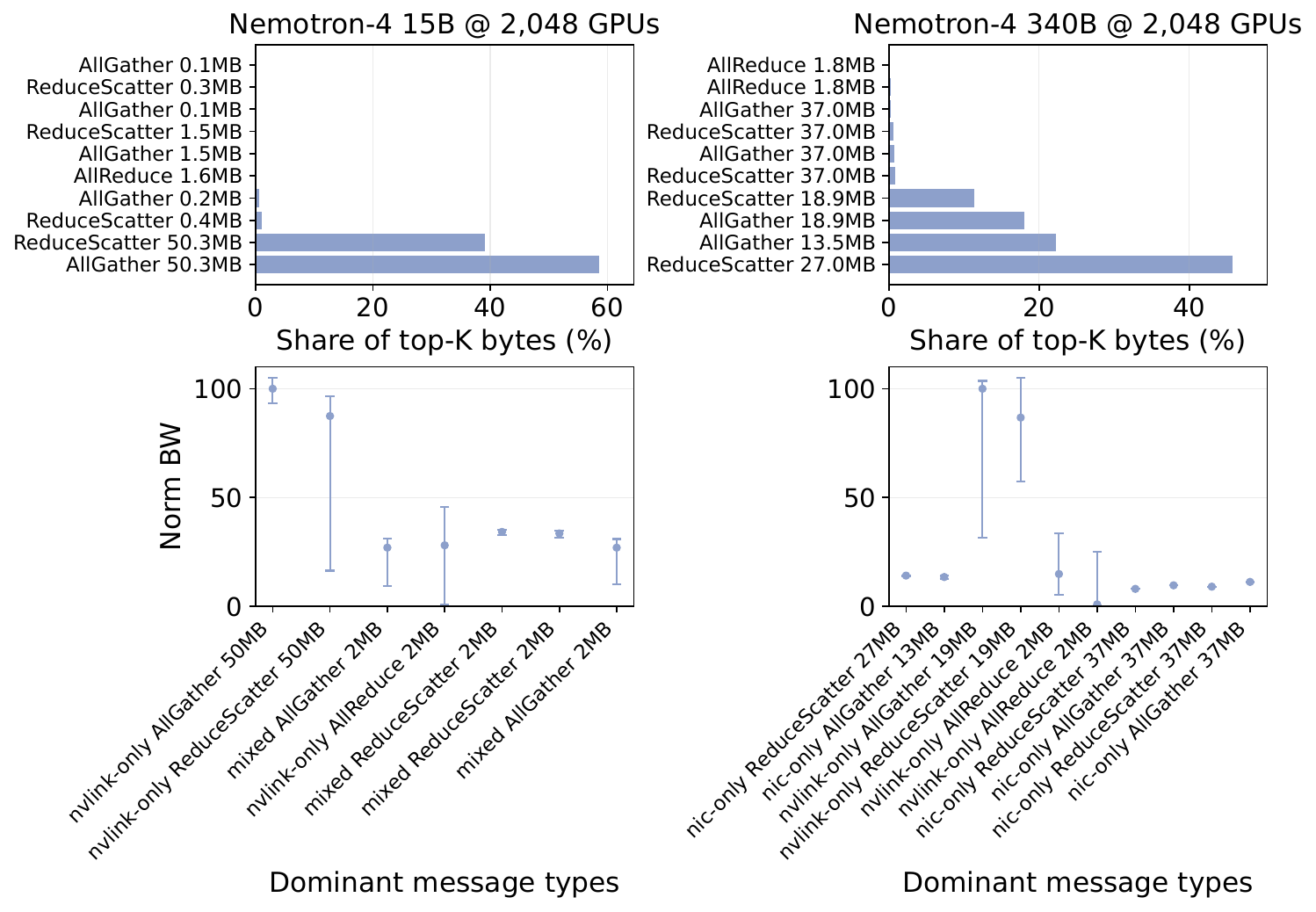}}
  \caption{Nemotron-4 at 2{,}048 GPUs: top row shows the top message sizes by total communicated bytes; bottom row shows bus-bandwidth percentiles (p10/p50/p90, normalized per panel) for frequent large messages ($\ge 1$ MB). Left column: 15B; right column: 340B.}
  \label{fig:340b_topmsgsize_and_bw}
\end{figure}

\subsection{Other Correlation Analysis}
\subsubsection{Observability into message-size distribution}
To demonstrate an example of other correlation analysis, we look at collective type, message size, and bus bandwidth for a Nemotron-4 15B at 16 GPUs.
Figure~\ref{fig:340b_topmsgsize_and_bw}(top) ranks the top-10 (collective, message-size) pairs by total communicated bytes, exposing that just a handful of recurring configurations account for the vast majority of traffic. Figure~\ref{fig:340b_topmsgsize_and_bw}(bottom) reports bus-bandwidth percentiles (p10/p50/p90) for those top-10 pairs, giving a view of both central tendency and variation.

\begin{tcolorbox}[colback=lightgreen!20!white,colframe=lightgreen!50!black,
                  left=2pt,right=2pt,top=2pt,bottom=2pt]
\textbf{Observation:} 
NVLink-only communicators exhibit the greatest amount of collective bandwidth variation while NIC-only communicators exhibit the least.
\end{tcolorbox}

We can use these observations, in combination with other analysis, such as variation attribution (in the next section), in order to guide co-design with ML frameworks to alleviate these potential performance fluctuations of collectives.

%% file: 5.NCCLtests.tex
\section{Case Study: Attributing Source of Collective Long-tail Variation}
\label{sec:nccl-tests-baseline}

From section~\ref{subsubsec:dominant-config-performance}, we observe significant variance in bandwidth of certain collectives. This case study aims to identify the root cause of such variation. Sources of bandwidth variation and long-tails in bandwidth can be due to hardware or due to the software stack, such as the ML framework coordinating the collective communication and ML workloads with many features such as concurrent compute/communication overlapping. 

\subsection{Attribution to ML Frameworks}

To isolate these effects, we ran two sets of experiments, once in a production Nemotron-4 340B training run (all prior experiments in this section), and one isolated in nccl-test. Again, we only look at the two most common communicator topology.  
Table~\ref{tab:nccl-tests-vs-training} reports the coefficient of variation (CV) of bus bandwidth for each scenario. We use CV rather than absolute standard deviation because absolute standard deviation scales with the mean and would bias comparison across regimes whose mean bandwidths span two orders of magnitude.



\begin{tcolorbox}[colback=lightgreen!20!white,colframe=lightgreen!50!black,
                  left=2pt,right=2pt,top=2pt,bottom=2pt]
\textbf{Observation:} 
Although mean collective bandwidth remains the same in nccl-tests and production training, the variation is significantly greater in production training.
\end{tcolorbox}

For both cases, the mean bus bandwidth in \texttt{nccl-tests} agrees with the production mean to within 4--8\%, but the variation is \emph{2 to 5 times larger in production training}: 0.336 versus 0.067 on the NVLink path and 0.174 versus 0.078 on the NIC path. This indicates that the ML framework that calls NCCL collectives is a potential source of collective performance variation. We speculate that this can be due to coordination of NCCL API calls or due to compute-communication overlapping of ML frameworks that can interfere with ongoing collectives.

\textbf{Limitation: overlap versus contention.} The production-versus-\texttt{nccl-tests} CV gap conflates two candidate mechanisms: coordination stalls introduced by the framework's scheduling of NCCL calls, and fabric contention from concurrent compute-communication overlap. The current data cannot separate the two, since doing so requires a controlled replay of the collective workload outside the training loop. As future work, we plan an instrumented replay of the recorded collective sequence (preserving communicator topologies, message sizes, and inter-arrival timing) together with annotation of overlap windows from framework hooks, which would attribute the excess variation to each mechanism separately.



\begin{table}[b!]
\centering
\footnotesize
\setlength{\tabcolsep}{3pt}
\caption{Bus bandwidth of the two dominant 340B@2{,}048 communicator topologies, measured in production training versus the same collective replayed in isolation under \texttt{nccl-tests} on the same cluster.  NVL = nvlink-only, NIC = NIC-only.}
\label{tab:nccl-tests-vs-training}
\begin{tabular}{@{}llrrr@{}}
\hline
Pick & Source & $n$ & Norm BW & CV \\
\hline
\multirow{2}{*}{NVL AllGather 8r/1n 18\,MB}
 & nccl-tests &      3{,}995 & 0.962 & \textbf{0.067} \\
 & training   &    551{,}982 & 1.000 & \textbf{0.336} \\
\hline
\multirow{2}{*}{NIC AllGather 32r/32n 13\,MB}
 & nccl-tests &     16{,}000 & 0.919 & \textbf{0.078} \\
 & training   &    955{,}137 & 1.000 & \textbf{0.174} \\
\hline
\end{tabular}
\end{table}

\subsection{Attribution to GPU Scaling}
To further investigate the cause of bandwidth variation, we now aim to eliminate the impact of GPU scaling on variance. 
Table~\ref{tab:scale-cv} reports the coefficient of variation(CV $=$ stddev/mean) of bus bandwidth for the dominant Nemotron-4 15B nvlink-only AllGather communicator ([2 ranks, 1 node], 50MB) across all five GPU scales.

\begin{table}[!t]
\centering
\small
\caption{Cross-scale dispersion of the dominant 15B nvlink-only AllGather bucket ([2,1], 50\,MiB).  
}
\label{tab:scale-cv}
\begin{tabular}{r r r r}
\hline
GPUs & $n$ ops & Norm BW & CV \\
\hline
16    &     600   & 0.989 & 0.041 \\
32    &  1{,}200  & 0.997 & 0.048 \\
64    &  9{,}000  & 1.000 & 0.046 \\
128   & 14{,}998  & 0.996 & 0.051 \\
2{,}048 & 643{,}178 & 0.993 & 0.056 \\
\hline
\end{tabular}
\end{table}

\begin{tcolorbox}[colback=lightgreen!20!white,colframe=lightgreen!50!black,
                  left=2pt,right=2pt,top=2pt,bottom=2pt]
\textbf{Observation:} Both the mean \emph{and} the coefficient of variation of bus bandwidth are stable as GPU count scales from 16 to 2{,}048. Therefore, relative variance of a fixed communicator topology is not dependent on GPU scaling.
\end{tcolorbox}

CV remains within a narrow band (0.041--0.056) across all GPU counts, with no meaningful increase, therefore, relative variation is not caused by GPU scaling, further justifying that variation is likely caused by the ML framework. Therefore, to improve collective performance and minimize collective variation, we require co-designing with ML frameworks. 

%% file: 6.Straggler.tex
\section{Case Study: Straggler Anomaly Localization and Attribution}
\label{sec:straggler-case}
In this case study, we demonstrate how NIXT's improved observability features can be used to localize a straggler anomaly in space and time and attribute it to the host resources of the job.
We compare three Nemotron-4 15B runs at 16 GPUs with identical configuration (TP=2). Two runs were scheduled on a host pool containing a known straggler, and one was collected after excluding that host. Because the workload is fixed, any bandwidth change reflects the host composition (allocated hardware) of the job rather than the collective itself.
Table~\ref{tab:straggler-cv} reports bus bandwidth for the dominant mixed AllGather bucket ([8 ranks, 2 nodes], 24.4\,MiB).

\begin{table}[b!]
\centering
\small
\caption{Nemotron-4 15B@16 GPUs, mixed AllGather [8r,2n,24.4\,MiB]. Mean bus bandwidth is normalized to the healthy baseline; CV is stddev divided by mean.}
\label{tab:straggler-cv}
\begin{tabular}{l r r r}
\hline
Run & $n$ ops & Norm BW & CV \\
\hline
Healthy      & 1{,}550  & 1.000 & 0.180 \\
Straggler-A  & 1{,}550  & 0.314 & 0.780 \\
Straggler-B  & 26{,}350 & 0.245 & 1.459 \\
\hline
\end{tabular}
\end{table}

Relative to the healthy baseline, both straggler runs reduce mean bus bandwidth by 3 to 4 times and increase CV by 4 to 8 times. The same pattern appears for the smaller intra-node bucket [8r,2n,18.9\,MiB], where mean bandwidth drops to 0.163 and 0.200 of healthy's normalized bandwidth. The effect is sustained across the dominant communicator topology and is therefore attributable to the host resources rather than to the collective.

\begin{tcolorbox}[colback=lightgreen!20!white,colframe=lightgreen!50!black,
                  left=2pt,right=2pt,top=2pt,bottom=2pt]
\textbf{Observation:} Compared to a healthy run, runs with stragglers show a 3 to 4 times drop in bandwidth and a 4 to 8 times increase in variability.
\end{tcolorbox}

\begin{figure}[t!]
  \centering
  \includegraphics[width=\linewidth]{\detokenize{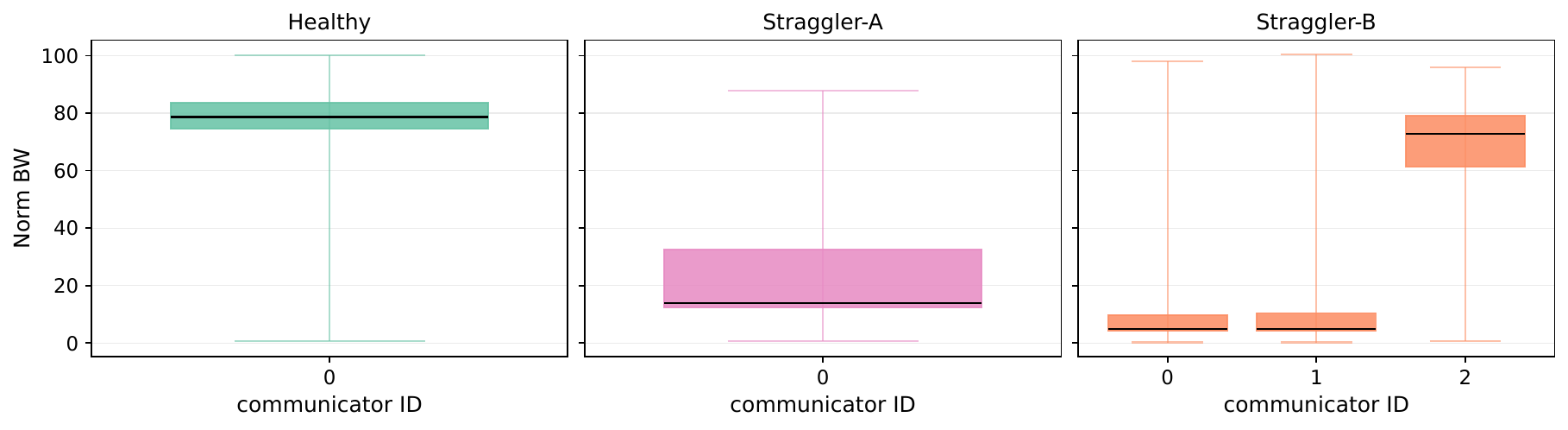}}
  \caption{Per-ID bandwidth for mixed AllGather 24\,MiB. Left: healthy. Right: straggler run.}
  \label{fig:straggler-perid-rangebar}
\end{figure}

To localize this regression across space and time, we summarize bus bandwidth by identifier and counter using sorted range bars in Figure~\ref{fig:straggler-perid-rangebar}. The whiskers show the minimum and maximum, the thick segment shows the interquartile range, and the marker shows the median. We apply this view to the straggler run and compare it with the healthy baseline on the dominant mixed AllGather communicator topology. 

\begin{figure*}[t!]
  \centering
  \includegraphics[width=\textwidth]{\detokenize{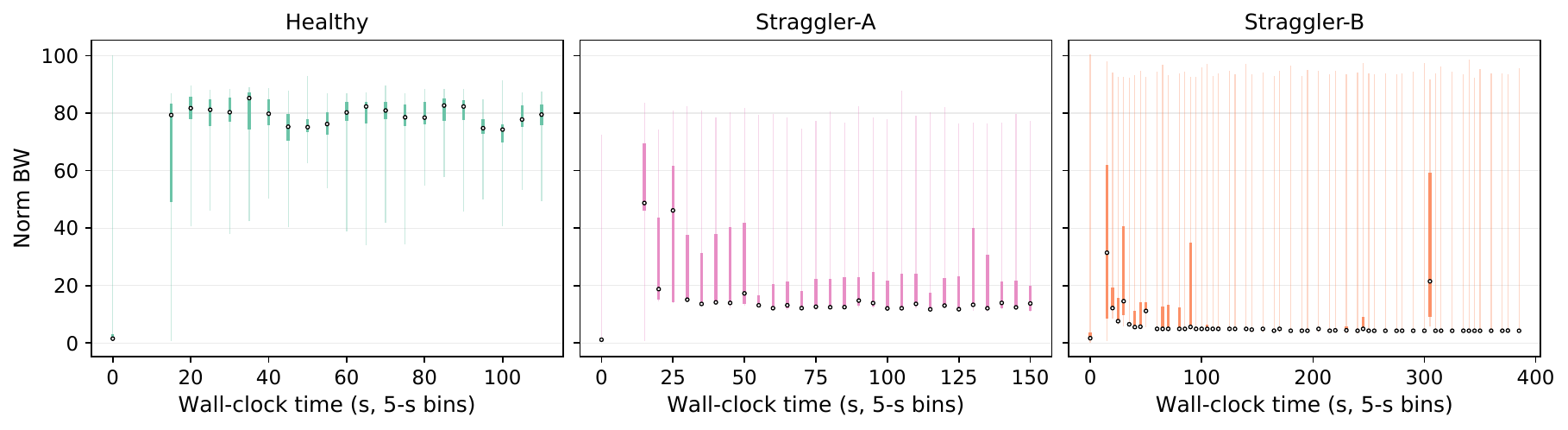}}
  \caption{Bandwidth versus time in 5\,s bins. Left: healthy. Right: straggler run.}
  \label{fig:straggler-pertimebin-rangebar}
\end{figure*}

\begin{tcolorbox}[colback=lightgreen!20!white,colframe=lightgreen!50!black,
                  left=2pt,right=2pt,top=2pt,bottom=2pt]
\textbf{Observation:} Identifier-based spatial correlation analysis can localize the straggler to a specific rank and identify when the bandwidth collapse is sustained. 
\end{tcolorbox}

Figure~\ref{fig:straggler-perid-rangebar} shows that the regression is host local. In the Straggler-B scenario, two of the three communicators associated with the dominant communicator topology collapse to about 10\,GB/s, while the remaining one stays close to the healthy median; the host-level view (omitted for space) mirrors this partition. Median CV rises from 0.18 in the healthy baseline to 1.33 per ID (1.38 per host). Without ID-level disaggregation, one would only identify a degradation in performance compared to Healthy, as in scenario Straggler-A. This analysis allows us to quickly localize the anomaly to the specific straggler hosts.

Figure~\ref{fig:straggler-pertimebin-rangebar} shows that the slowdown is sustained and not transient. Once the affected hosts initiate the first few collectives, median bandwidth remains near 10\,GB/s for the full test window; the per-communicator sequence-number axis (omitted for space) shows the same flat-and-low profile. In contrast, the healthy baseline reaches about 170\,GB/s within the first 20\,s and remains stable.

\textbf{Scope and limitation.} This case study performs \emph{anomaly localization and attribution}: NIXT isolates the performance signature of the degradation --- bandwidth collapse, asymmetry across ranks and hosts, and sustained variance --- and attributes it to the host composition of the job. NIXT does not directly probe the fabric: link-level health indicators such as NVLink CRC errors and InfiniBand error counters are reported by DCGM and switch telemetry, and NIXT is designed to correlate with those sources rather than replace them. Confirming the underlying root cause (e.g., a degraded link, thermal throttling, or a failing device) by joining NIXT's localization output with DCGM, thermal, and link-error telemetry is the natural next step, which we leave as future work.

%% file: 7.Related.tex
\section{Related Works}
\label{sec:related}


\paragraph{Performance observability for GPU clusters.} Cluster-scale observability traditionally combines host- and GPU-level metric exporters such as Node Exporter~\cite{node_exporter} and DCGM Exporter~\cite{nvidia_dcgm_exporter} with Prometheus~\cite{prometheus_overview_2025} and OpenTelemetry pipelines~\cite{opentelemetry_ai_best_practices}. 
For NCCL specifically, the PyTorch Flight Recorder~\cite{pytorch_flight_recorder2024} captures recent collectives on failure for post-mortem debugging, and eACGM~\cite{xu2025eacgm} uses eBPF~\cite{eBPF} to trace GPU events without source instrumentation. 
A parallel line of work builds low-overhead tracing for production systems at large: Hindsight~\cite{hindsight2023} retains detailed traces only when an anomaly materializes; DeepFlow~\cite{deepflow2023} and TraceWeaver~\cite{traceweaver2024} reconstruct request paths without application modification; and DeepTrace~\cite{deeptrace2025}, EXIST~\cite{exist2025}, and Mint~\cite{mint2025} drive always-on tracing overhead low enough for hyperscale deployment, while Loom~\cite{loom2025} captures and queries high-frequency telemetry streams. 
On the accelerator side, Neutrino~\cite{neutrino2025} probes inside GPU kernels via programmable instrumentation, DeepContext~\cite{deepcontext2025} profiles deep-learning workloads across frameworks, DFTracer~\cite{dftracer2024} and TraceFlow~\cite{traceflow2025} address AI-workflow and large-scale parallel trace analysis, and Huawei describes an industrial profiling and analysis system for Ascend training clusters~\cite{zhou2025ascend}. 
These systems trace RPCs, kernels, or workflow I/O; none expose the semantics of collective communication. NIXT brings the same always-on, low-overhead posture to the NCCL layer itself.

\paragraph{Reliability and diagnosis in large-scale training.}
Experience and characterization studies from ByteDance~\cite{jiang2024megascale,wan2025robust}, Shanghai AI Lab~\cite{hu2024characterization}, and Meta~\cite{li2024revisiting} document how often collective communication is implicated in failures and slowdowns of production LLM training. Production diagnosis systems localize faults from different signals: Minder~\cite{deng2025minder} compares host metrics across data-parallel replicas, Holmes~\cite{yao2025holmes} compares iteration-level profiles across ranks, FLARE~\cite{316592} diagnoses divergent training runs, Aegis~\cite{dong2025aegis} operates as a long-running fault-diagnosis service, and SuperBench~\cite{xiong2024superbench} proactively validates infrastructure with synthetic benchmarks. 
A complementary thread targets failures that never crash: fleet-scale studies characterize silent data corruptions in production CPUs~\cite{sdc2023cpu}, Dr.~DNA~\cite{drdna2024} and automated proactive checks~\cite{trainingconfidence2025} catch silent errors in deep-learning training, semantic checkers derived from tests detect silent failures in distributed systems~\cite{semanticcheckers2025}, and slow-fault studies show how poorly current systems tolerate degraded-but-alive components~\cite{onesize2025}; recent work even applies LLMs to root-cause cloud incidents~\cite{llmrca2024}. 
Network-side monitors such as R-Pingmesh~\cite{liu2024rpingmesh} and Hawkeye~\cite{wang2025hawkeye} diagnose the RoCE/RDMA fabric underneath the collective layer. NIXT is complementary infrastructure to these diagnosers: an exporter substrate at the NCCL layer whose relational output they can consume.

\paragraph{Stragglers and anomaly detection.}
Straggler effects in distributed training have been studied through what-if analysis~\cite{308748}, malleable parallelization~\cite{li2024malleus}, communication--computation trade-offs~\cite{ozfatura2020straggler}, tail-tolerant collective designs~\cite{optireduce2025}, and dependency tracing within NCCL~\cite{deng2025mycroft}; deadlock-prevention work~\cite{deadlock2025} attacks preventively the same collective hangs that tracing systems detect reactively. In adjacent domains, FBDetect~\cite{yoon2024fbdetect} catches sub-percent regressions in production via in-production monitoring.

\paragraph{Optimizing collective communication.}
An orthogonal body of work reshapes the collectives themselves: TACCL~\cite{taccl2023} and MSCCLang~\cite{mscclang2023} synthesize and program custom collective algorithms, MCCS~\cite{mccs2024} lifts collectives into a managed cloud service, AutoCCL~\cite{autoccl2025} tunes communication parameters online, T3~\cite{t3overlap2024} transparently overlaps collectives with computation, and TCCL~\cite{tccl2024} discovers better communication paths on PCIe clusters. These optimizers change exactly the behavior NIXT records, and the measurements NIXT exports for every collective on each rank provide the ground truth such systems need for tuning and validation.

NCCL  Inspector~\cite{nvidia_nccl_inspector,nccl_inspector_guide,nvidia_inspector_observability_2025,nvidia_inspector_prometheus_2025,nvidia_nccl_observability_2024} provides continuous profiling of individual collective calls at the NCCL layer.

NIXT builds on Inspector's raw output and supplies the missing analysis layer: a relational schema, dispersion-focused queries, and the Identifier/Counter/Measurement decomposition that lets operators correlate collective performance across the rank-time grid.

%% file: Conclusion.tex
\section{Conclusion}
\label{sec:conclusion}

This paper introduces NIXT, a \underline{N}CCL \underline{I}nspector E\underline{x}porter \underline{T}ool for NCCL Inspector. NIXT tackles the problem of high volume and dimension of raw NCCL Inspector output in order to provide actionable insights into collective performance characteristics and accelerate root cause analysis and anomaly detection.


Evaluation across H100 GPU clusters validates that NIXT provides unprecedented visibility into collective communication operations with minimal performance impact, making it suitable for production deployment at enterprise scale.
Using Nemotron-4 pretraining workloads on up to 2{,}048 H100 GPUs, we showed that NIXT reveals communication structure that was previously opaque to conventional monitoring. Our case study highlighted that a small number of communicator topologies and message-size regimes dominate collective traffic, that the bus bandwidth of their individual calls remains largely stable as model and cluster scales increase, and that spatial and temporal correlation analyses can distinguish healthy, regular communication motifs from configurations with high variability. We further show how NIXT can localize GPU straggler anomalies, attribute them to host resources, and pinpoint potential sources of collective performance variation.

While our case studies focus on LLM pretraining, the taxonomy and analysis primitives of NIXT are workload-agnostic: any application that produces NCCL Inspector logs --- reinforcement-learning rollouts, inference serving, HPC collectives, or \texttt{nccl-tests} --- can be analyzed with the same primitives. NIXT is deployed in production clusters, and we plan to report on additional model families and cluster configurations in future work.

Our work addresses the critical gap in communication-layer observability for modern GPU clusters, where traditional monitoring approaches fail to capture the complex collective operation patterns fundamental to distributed training. The integration with established monitoring infrastructure ensures practical adoption in existing operational environments.



%% file: artifact_appendix.tex
\section*{Artifact Appendix}

\subsection*{Abstract}

This artifact contains the complete source code of NIXT (NCCL
Inspector eXporter Tool), the observability pipeline used in the
paper: (1)~the \textbf{Inspector} NCCL profiler plugin (C++), which
attaches to any NCCL-based job and emits per-rank, per-collective
JSON dumps; (2)~the \textbf{exporter} (Python), which parses the
dumps into parquet tables and summary reports; and (3)~the
\textbf{analysis} scripts (Python) that generated every experiment
figure in the paper.

The telemetry analyzed in the paper was collected from large-scale
production LLM training runs (up to 2048 H100 GPUs) whose data is
confidential and cannot be redistributed. The artifact is therefore
\emph{code only}: we claim the \textbf{Available} and
\textbf{Reviewed} badges, not \textbf{Reproducible}. To let
evaluators verify functionality end-to-end, the artifact includes a
single-script demo that runs the full pipeline (plugin
$\rightarrow$ dumps $\rightarrow$ parquet $\rightarrow$ summaries)
on a small GPU node using \texttt{nccl-tests} as the workload, plus
reference copies of all paper figures in \texttt{expected\_output/}
and a figure-to-script map in \texttt{docs/figure\_map.md}.

\subsection*{Artifact check-list (meta-information)}

{\small
\begin{itemize}
  \item {\bf Program: } NCCL Inspector profiler plugin (C++);
        exporter and analysis scripts (Python);
        \texttt{nccl-tests} (fetched automatically by the demo).
  \item {\bf Compilation: } GNU make, \texttt{g++} (C++14/17),
        CUDA toolkit $\geq$ 12.x.
  \item {\bf Binary: } \texttt{libnccl-profiler-inspector.so},
        built from source.
  \item {\bf Data set: } none distributed (production telemetry is
        confidential); the demo generates its own data via
        \texttt{nccl-tests}.
  \item {\bf Run-time environment: } Linux x86\_64; CUDA $\geq$ 12.x;
        NCCL $\geq$ 2.28 (2.28--2.30 tested); Python $\geq$ 3.10.
  \item {\bf Hardware: } one node with $\geq$ 2 NVIDIA GPUs
        (developed on A100/H100; no architecture-specific code).
  \item {\bf Execution: } single script \texttt{demo/run\_demo.sh}.
  \item {\bf Metrics: } per-collective bus bandwidth, execution
        time, message sizes, and transferred bytes.
  \item {\bf Output: } per-rank \texttt{*.log.gz} JSON dumps;
        parquet tables; summary CSVs and plots; reference paper
        figures in \texttt{expected\_output/}.
  \item {\bf Experiments: } end-to-end functional demo; paper
        analysis scripts provided for inspection and reuse.
  \item {\bf How much disk space required (approximately)?: }
        $\sim$2 GB (dominated by the \texttt{nccl-tests} build).
  \item {\bf How much time is needed to prepare workflow
        (approximately)?: } $\sim$15 minutes.
  \item {\bf How much time is needed to complete experiments
        (approximately)?: } $\sim$15 minutes.
  \item {\bf Publicly available?: } Yes (GitHub; archived on
        Zenodo).
  \item {\bf Code licenses (if publicly available)?: }
        BSD-3-Clause (Inspector plugin vendored from NVIDIA's NCCL
        \texttt{ext-profiler} example, NVIDIA copyright).
  \item {\bf Archived (provide DOI)?: }
        \texttt{10.5281/zenodo.21755872}.
\end{itemize}
}

\subsection*{Description}

\subsubsection*{How to access}

The artifact is available at
\url{https://github.com/ziyang-arch/nixt-analysis}
and archived at DOI \texttt{10.5281/zenodo.21755872}.
Repository layout:

{\small
\begin{itemize}
  \item \texttt{inspector/} --- NCCL Inspector profiler plugin
        (C++, standalone Makefile).
  \item \texttt{exporter/} ---
        \texttt{perf\_summary\_exporter.py}: parses dumps into
        parquet + summary reports/plots.
  \item \texttt{analysis/} --- paper figure scripts.
  \item \texttt{demo/} --- end-to-end functional demo (2--8 GPUs).
  \item \texttt{expected\_output/} --- experiment figures exactly
        as they appear in the paper.
  \item \texttt{docs/figure\_map.md} --- paper figure
        $\leftrightarrow$ script $\leftrightarrow$ input mapping.
\end{itemize}
}

\subsubsection*{Hardware dependencies}

A Linux x86\_64 node with at least 2 NVIDIA GPUs. Any recent
architecture works; the demo was developed against A100/H100.

\subsubsection*{Software dependencies}

CUDA $\geq$ 12.x; NCCL $\geq$ 2.28 (profiler plugin API; 2.28--2.30
tested); Python $\geq$ 3.10 with \texttt{pandas}, \texttt{tqdm},
\texttt{duckdb}, \texttt{matplotlib}, \texttt{pyarrow},
\texttt{numpy} (\texttt{pip install -r exporter/requirements.txt});
\texttt{nccl-tests} (cloned and built automatically by the demo).

\subsubsection*{Data sets}

The paper's telemetry comes from confidential production training
runs and is not redistributable. The demo generates its own
Inspector dumps from \texttt{nccl-tests} collectives.

\subsection*{Installation}

{\small
\begin{verbatim}
pip install -r exporter/requirements.txt
make -C inspector   # set CUDA_HOME if CUDA is
                    # not at /usr/local/cuda
\end{verbatim}
}

\subsection*{Experiment workflow}

The pipeline is: Inspector plugin (attached to a NCCL job via
\texttt{NCCL\_PROFILER\_PLUGIN}) $\rightarrow$ per-rank
\texttt{*.log.gz} JSON dumps $\rightarrow$ exporter $\rightarrow$
parquet + summaries $\rightarrow$ analysis scripts $\rightarrow$
figures. The demo automates all stages:

{\small
\begin{verbatim}
NGPUS=2 ./demo/run_demo.sh
\end{verbatim}
}

It (1)~builds the plugin, (2)~clones/builds \texttt{nccl-tests} if
needed, (3)~runs \texttt{all\_reduce\_perf} and
\texttt{all\_gather\_perf} (1\,KB--256\,MB) with the plugin
attached, and (4)~exports the dumps to parquet and summary
reports.

\subsection*{Evaluation and expected results}

After the demo completes ($\sim$15 min total), verify:

{\small
\begin{itemize}
  \item \texttt{demo/dump/<timestamp>/} contains per-rank
        \texttt{*.log.gz} Inspector dumps (the script fails loudly
        if none are produced);
  \item \texttt{data/demo-analysis/parquet\_files/} contains one
        parquet file per rank;
  \item \texttt{data/demo-analysis/} contains summary CSVs and
        plots with plausible bus-bandwidth and message-size
        statistics for the two collectives.
\end{itemize}
}

This validates the full plugin $\rightarrow$ exporter path used for
every result in the paper. The scripts in \texttt{analysis/}
generated all experiment figures; \texttt{docs/figure\_map.md} maps
each paper figure to its script, and \texttt{expected\_output/}
contains the exact figures from the camera-ready paper. Because the
underlying telemetry is confidential, re-generating those figures
is out of scope for this evaluation (we do not claim the
\textbf{Reproducible} badge).

\subsection*{Experiment customization}

The plugin attaches to any NCCL/PyTorch job:

{\small
\begin{verbatim}
export NCCL_PROFILER_PLUGIN=\
  /path/to/libnccl-profiler-inspector.so
export NCCL_INSPECTOR_ENABLE=1
export NCCL_INSPECTOR_DUMP_DIR=/path/to/dumps
<launch your job as usual>
python3 exporter/perf_summary_exporter.py \
  --input_dir /path/to/dumps
\end{verbatim}
}

All analysis scripts resolve inputs/outputs relative to
\texttt{NIXT\_ROOT} (defaults to the repository root), so they can
be pointed at telemetry collected from the evaluator's own
workloads. \texttt{inspector/README.md} documents all knobs
(verbose event traces, Prometheus export, dump intervals).

\subsection*{Methodology}

Submission, reviewing and badging methodology:

\begin{itemize}
  \item \url{https://www.acm.org/publications/policies/artifact-review-and-badging-current}
  \item \url{https://cTuning.org/ae}
\end{itemize}